\documentclass[%
amsmath,amssymb,
aps,
prb,
twocolumn
]{revtex4-2}

\usepackage{graphicx}
\usepackage{dcolumn}
\usepackage{bm}
\usepackage{xcolor}
\usepackage[caption=false]{subfig}
\usepackage{tikz}
\usepackage{xfrac}
\usepackage{blindtext}
\usepackage{times}

\definecolor{darkblue}{HTML}{004D6B}
\definecolor{darkred}{HTML}{8c1515}

\usepackage{hyperref}
\hypersetup{
	colorlinks=true,
	urlcolor=darkred,
	citecolor=darkblue,
	linkcolor=darkred,
	breaklinks
}
\usepackage{physics}
\usepackage{bbold} 

\begin{document}
	
	\title{Stability of Floquet Majorana box qubits}
	
	\author{Anne Matthies$^1$}
	\author{Jinhong Park$^1$}
	\author{Erez Berg$^2$}
	\author{Achim Rosch$^1$}
	\affiliation{$^1$ Institute for Theoretical Physics, University of Cologne, 50937 Cologne, Germany, $^2$Department of Condensed Matter Physics, Weizmann Institute of Science, Rehovot, Israel.}
	
	\date{\today}
	
	\begin{abstract}
		In one-dimensional topological superconductors driven periodically with the frequency $\omega$, two types of topological edge modes may appear, the well-known Majorana zero mode and a Floquet Majorana mode located at the quasi-energy $\hbar \omega/2$. We investigate two Josephson-coupled topological quantum wires in the presence of Coulomb interactions, forming a so-called Majorana box qubit. An oscillating gate voltage can induce Floquet Majorana modes in both wires. This allows encoding 3 qubits in a sector with fixed electron parity. If such a system is prepared by  increasing the amplitude of oscillations adiabatically, it is inherently unstable as interactions resonantly create quasi particles. This can be avoided by using instead a protocol where the oscillation frequency is increased slowly. In this case, one can find a parameter regime where the system
		remains stable.
	\end{abstract}
	
	\maketitle


	Majorana zero modes (MZMs) arise at the edges of one-dimensional topological superconductor (TS) \cite{Kitaev2001}. 
	A pair of MZMs can encode a qubit. MZMs allow to store quantum information non-locally, and are thus protected against local noise. In combination with their non-Abelian braiding statistics~\cite{Ivanov2001,Read2000}, MZMs are  promising candidates for topological quantum computation. Such systems can be realized in different physical setups, 
	such as semiconducting quantum wires with strong spin-orbit coupling or from topological insulators, proximity coupled to conventional superconductors \cite{Fu2008,Alicea2010,Oreg2010,Lutchyn2010,Brouwer2011,Manousakis2017,Lutchyn2018,Flensberg2021}.

\begin{figure}[b]
		\includegraphics[width=0.75 \columnwidth]{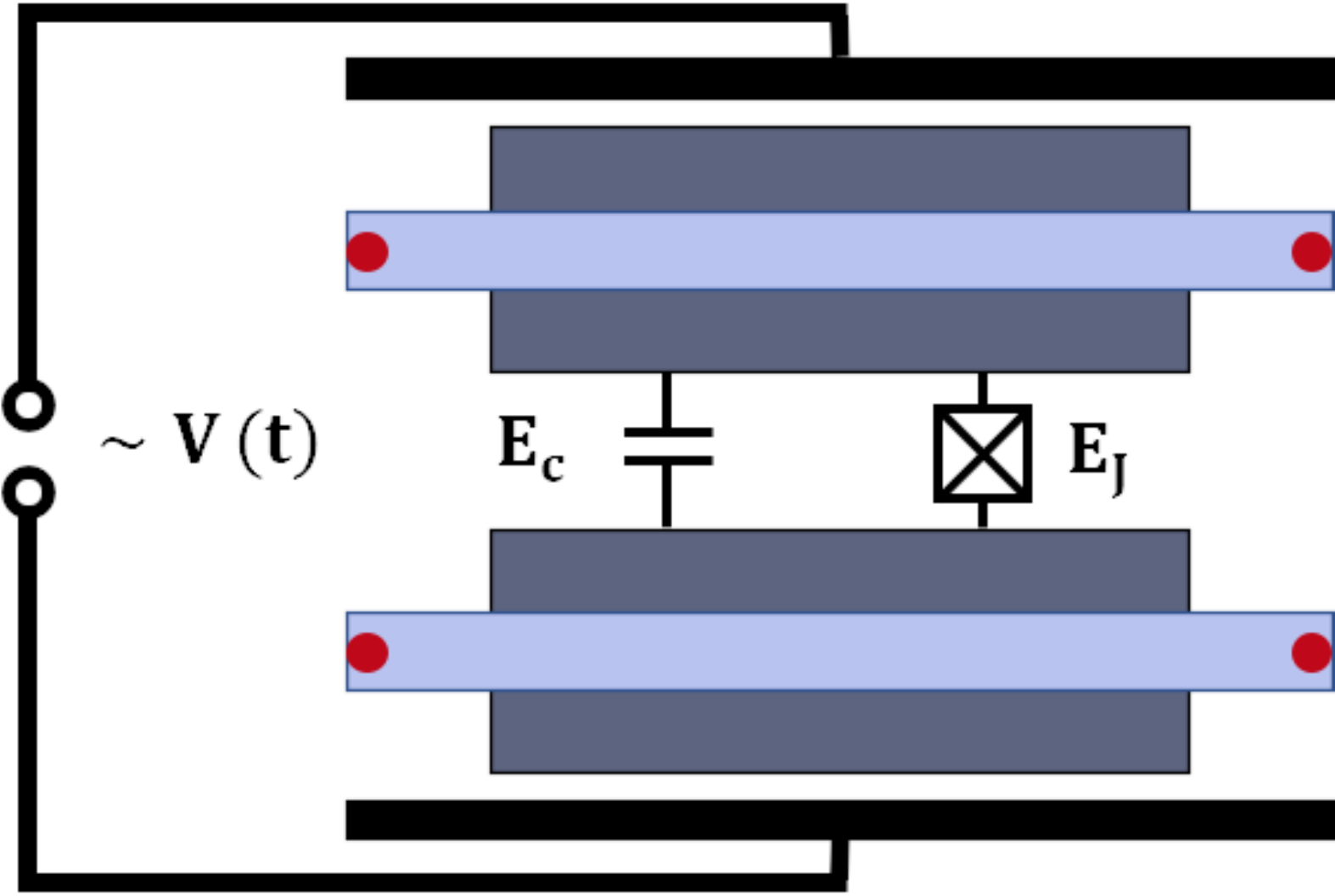}
		\caption{\label{fig:FloquetBox} Schematic of the Floquet Majorana Box Qubit: Two topological superconducting nanowires (light blue) proximity coupled to two bulk superconductors (dark blue).The bulk superconductors are Josephson coupled with energy $E_J$ and in the Coulomb blockage regime with large $E_c$. The bulk superconductors are driven by a capacitively coupled voltage $V(\tau)$.
		}
	\end{figure}

	Two Majoranas at the ends of a single quantum wire, however, do not define a useful qubit. In this case, the quantum information is encoded in the electron number parity of the system. It is very difficult to create coherent superpositions of such states and, furthermore,  if the system is not grounded, the long-ranged Coulomb interaction
	leads to an energetic splitting between the two parity states. 
	Therefore, a different setup - the Majorana box qubit - has been introduced \cite{Plugge2017,Karzig2017,Vijay2016,Flensberg2021}. As shown in Fig.~\ref{fig:FloquetBox}, two topological superconducting quantum wires on a floating superconducting island are coupled by Josephson junctions. Charging one wire relative to the other costs a finite energy. Due to this charging energy, the four-dimensional low-energy Hilbert space splits into two  sectors with even and odd parity. Thus one obtains a single qubit, which can be embedded in larger quantum circuits \cite{Karzig2017}.
	
	In this paper, we combine the idea of a Majorana box qubit with another powerful concept. In periodically driven systems, one can realize topological phases with no equilibrium equivalent \cite{Kitagawa2010,Wilczek2012,Rudner2013,nathan2015topological,Keyserlingk2016,Else2016,Else2016a,Potter2016,Titum2016,Roy2016,Khemani2016a}. It has been shown that driving a single topological superconductor periodically
	leads to the emergence of an extra pair of topologically protected Majorana modes \cite{Jiang2011,Liu2012,Kundu2013,Li2014,Liu2019,Peng2021} or even more Majorana states, when multiple incommensurate driving frequencies are used \cite{Peng2018}. Similarly to MZMs in static systems, Floquet Majorana modes, also called Majorana $\pi$ modes (MPM), are localized to the ends of the system. However, their quasi-energy is $\pi \frac{\hbar}{T}=\frac{\hbar \omega}{2}$, where $\omega=\frac{2 \pi}{T}$ is the driving frequency and $T$ the period of driving. As Floquet Majoranas and ordinary MZMs have different quasi-energies, they do not hybridize. 
	
	This allows for a topologically protected braiding operation \cite{Bauer2019} even in a one-dimensional wire, given that the time-dependent Hamiltonian is nearly perfectly periodic in time.
	
	We will show how this principle can be used to construct a ``Floquet Majorana box'' by simply driving the system with an AC voltage which is applied via gates
	to the two superconducting islands as shown in Fig.~\ref{fig:FloquetBox}. 
	
	The device hosts a total of three logical qubits in each subspace of fixed electron parity. A similar device based on the corner modes of a second-order topological
	superconductor was recently considered by Bomantara and Gong
	\cite{Bomantara2020}. They showed that such a setup allows for all Clifford gate operations, both for one and two qubits, in a topologically protected way and can be used
	for braiding experiments.
	
	The main goal of our study is to explore interaction effects, which are unavoidable in a device based on Coulomb blockade. 
	We will first derive an effective low-energy theory for the Floquet Majorana box qubit which allows us to study the emergence of Floquet Majoranas.
	Using perturbation theory, we explore interaction effects and show that the stability of the system depends decisively on the preparation protocol
	of the driven system. 
	
	{\it Model:} To obtain a minimal model for the physics of the device shown in Fig.~\ref{fig:FloquetBox}, we consider two Kitaev chains, $\alpha=1,2$, coupled to two superconducting islands. 
	\begin{widetext}
		\begin{equation}
			H=\frac{E_c}{2} \hat{Q}^2 - E_J \cos( \hat{\phi})+ \hat{Q} V(\tau)-
			\sum_{i,\alpha} \mu c_{i,\alpha}^{\dagger}c_{i,\alpha}+\sum_{i,\alpha} \left(-\frac{t}{2}c_{i,\alpha}^\dagger c_{i+1,\alpha}+\frac{\Delta}{2}e^{(-1)^\alpha i \hat{\phi}/2} c_{i,\alpha}^\dagger c_{i+1,\alpha}^\dagger+h.c.\right)\label{eq:H}
		\end{equation}
	\end{widetext}
	Here $\hat \phi=\hat\phi_1-\hat \phi_2 $ is the phase difference of the two superconducting islands and the operator 
	$\hat Q=\hat Q_1-\hat Q_2$ counts the difference in the number of cooper pairs on the two islands. $E_c$ is the charging energy associated with this charge imbalance and $E_J$ denotes the Josephson coupling arising from Cooper pair tunneling, 
	with $[\hat{Q},e^{\pm i \hat{\phi}/2}]=\pm    e^{\pm i \hat{\phi}/2}$. The system is driven by a time-dependent ac voltage,
	$V(\tau)=V_0 \cos(\omega \tau)$ where we denote time by $\tau$. For simplicity, we model the interactions here by the charging term $\frac{E_c}{2} \hat{Q}^2$ only (omitting further interactions both within each Majorana chain and between the chains and the superconducting islands), but we will argue below that all qualitative results remain valid when further weak interactions are included.
	Furthermore, the global charging energy $Q_{\rm{tot}}^2/(2 C)$ has been omitted as we consider only processes where the total charge $Q_{\rm{tot}}$ is conserved.

	We assume that all energy scales associated with the quantum wires and the driving frequency $\omega$
	are much smaller than the 
	driving amplitude
	$V_0$ and the energy scales associated with the large
	superconducting islands, $\omega, \Delta, t, \mu \ll E_c\ll E_J, V_0$. The limit $E_c \ll E_J$ is  necessary  for the operation of the box qubit, exponentially suppressing the relative charging energy between the quantum wires
	and the sensitivity to charge noise \cite{Koch2007, Karzig2017}.
	Since $E_c \ll E_J$, phase fluctuations are small, $\cos \hat\phi \approx 1-\frac{\hat\phi^2}{2}$, and the superconducting islands are well described by a time-dependent harmonic oscillator
	\begin{equation}
		H_{SC}(\tau)=\frac{E_c}{2} \hat{Q}^2+\frac{E_J}{2} \hat{\phi}^2+\hat{Q} V(\tau).
	\end{equation}
	To derive an effective low-energy Hamiltonian for the driven Kitaev chains, we apply three unitary transformations  (see Appendix \ref{app:A} for details):  We shift the $\hat Q$ coordinate of the  harmonic oscillator to absorb the linear term $\hat Q V(\tau)$ by the unitary  $U_1(\tau)=e^{-i \hat \phi V(\tau)/(2E_c)}$. The resulting time-dependence of $\hat Q$ induces a time dependent phase, $\hat \phi \sim \partial_\tau Q /E_J$
	, which we eliminate  by a second transformation $U_2=e^{-i \hat Q \partial_\tau V(\tau)/(4 E_J E_c)}$. Finally, we use a Schrieffer-Wolff transformation $U_3=e^{-i\hat{Q}\frac{\Delta}{8E_J} T }$ to eliminate perturbatively transitions to excited states of $H_{SC}$ triggered by $e^{\pm i \hat \phi/2}$ in Eq.~\eqref{eq:H}.
	All approximations are well-controlled for $\Delta \ll E_J$, $E_c \ll E_J$, and $V_0 \ll \frac{E_c E_J}{\omega}$, 
	see App.~\ref{app:A}. 
	We thus arrive at the time-dependent  low-energy Hamiltonian 
	\begin{widetext}
		\begin{align}
			H_{\rm eff}(\tau)&=-\sum_{i,\alpha} \mu  c^\dagger_{i,\alpha} c_{i,\alpha}+\sum_{i,\alpha}\left(- \frac{t}{2} c^\dagger_{i,\alpha} c_{i+1,\alpha} +\frac{\Delta}{2} e^{-(-1)^\alpha i\frac{\dot{V}(\tau)}{2E_c E_J}} c^\dagger_{i,\alpha} c^\dagger_{i+1,\alpha} +h.c.\right)+\frac{\Delta^2}{32E_J}T^2, 
			\label{eq:Heff1} \\
			T&=i e^{-i(-1)^\alpha\frac{\dot{V}(\tau)}{2E_c E_J}}\sum_{i,\alpha}  c^\dagger_{i,\alpha} c^\dagger_{i+1,\alpha}+h.c. \nonumber
		\end{align}
	\end{widetext}

	\begin{figure}
		\includegraphics[width=0.99 \columnwidth]{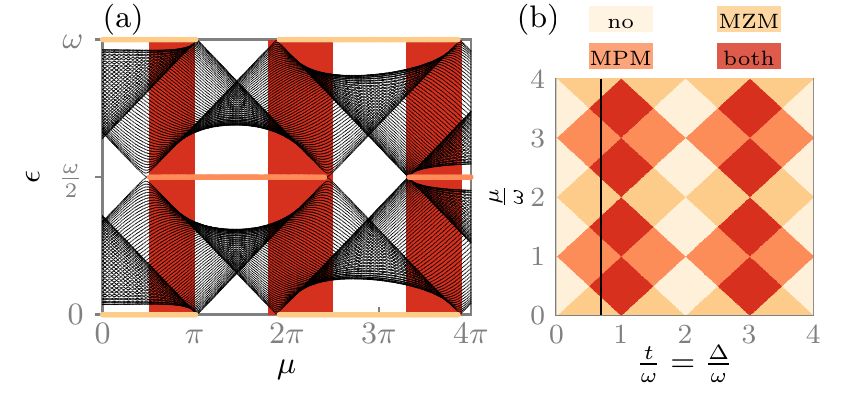}
		\caption{\label{fig:floquetKitaevChain} (a)
			Floquet spectrum	as a function of the static chemical potential $\mu$ ($N=60$, $\delta\mu=3\pi$, $t=\pi$, $T=1.4$, $N_{\text{steps}}=100$, $\Delta=-\pi$). The Majorana zero mode (MZM) and the Majorana $\pi$ modes (MPM) are depicted  in light orange and orange, respectively. Areas of co-existing MZM and MPM are shaded in red. (b) Phase diagram of the driven Kitaev chain as function of $\frac{t}{\omega}=\frac{\Delta}{\omega}$ and $\frac{\mu}{\omega}$. It shows the four different phases: trivial, only MZM, only MPM and both modes.}
	\end{figure}
	
	We first study the non-interacting part of $H_{\rm eff}$ considering the small interaction effects encoded in the last term of Eq.~\eqref{eq:Heff1} only later. The time-dependent phase can be absorbed by a gauge transformation and rewritten as a time-dependent chemical potential
	\begin{align} \label{EffectiveHamiltonian}
		H^0_{\text{eff}}=&\sum_{i,\alpha}-\left(\mu-(-1)^\alpha \delta \mu\, \cos(\omega \tau)\right) c^\dagger_{i,\alpha} c_{i,\alpha} \nonumber\\
		&\quad -\frac{t}{2} c^\dagger_{i,\alpha} c_{i+1,\alpha}+\frac{\Delta}{2}c^\dagger_{i,\alpha} c^\dagger_{i+1,\alpha}+h.c.
	\end{align}
	
	with 	$\delta \mu = V_0 \frac{\omega^2}{E_c E_J}$
	for an oscillating gate voltage, $V(\tau)=V_0 \cos(\omega \tau)$.	
	The spectrum and the topological Floquet phases of this non-interacting Hamiltonian, which have been studied before 
	\cite{Bauer2019,Jiang2011,Liu2012}, are shown in  Fig.~\ref{fig:floquetKitaevChain}. Since energy is only conserved modulo the driving frequency $\omega$, this so called quasi-energy is shown in the Floquet spectrum in Fig.~\ref{fig:floquetKitaevChain}. Importantly, one obtains phases (red shaded regions in 
	Fig.~\ref{fig:floquetKitaevChain}a) where  Majorana zero modes and Floquet Majorana modes at quasi-energy $\pi/T=\omega/2$ coexist.	The spectrum was  obtained numerically by calculating the Floquet time evolution operator $U_F$ for a finite-size system over one period  $T$  by a Suzuki-Trotter decomposition using the Floquet Hamiltonian $H_F=i \log(U_F)/T$. Equivalently, the spectrum can be obtained by solving the Floquet matrix, see App.~\ref{app:bulkCalc}.
	The phase diagram of Fig.~\ref{fig:floquetKitaevChain} is obtained from the topological charges of the bulk bands, see \cite{Jiang2011,Bauer2019}. 
	
	Importantly, the Floquet spectrum is {\em not}  sufficient to describe the system, 
	as it does not specify the many-body (Floquet-) state, which may depend on the way the system is prepared.
	We first consider the standard setup for adiabatic state preparation by assuming that system is initially in the ground state of the Hamiltonian~\eqref{eq:H} with $V_0=0$. The oscillating gate voltage is then switched on slowly. In the following, we will focus on the bulk problem but we have checked numerically that the same physics also governs finite-size system.
	The initial state (with $V(\tau)=0$ )
	is the ground state $|GS\rangle$, and the excited states $a^\dagger_k|GS\rangle$
	have {\em positive} energies, $E_k\ge 0$, while $a_k|GS\rangle=0$. Within the Bogoliubov theory, one formally introduces {\em negative} energy states using $E_{k} a_{k}^\dagger a_{k}=-E_{k}  a_{k} a^\dagger_k$. We call these negative energies $-E_k$ `Bogoliubov shadows' to emphasize that the physical excitations have positive energies.
	
\begin{figure}
		\includegraphics[width=\columnwidth]{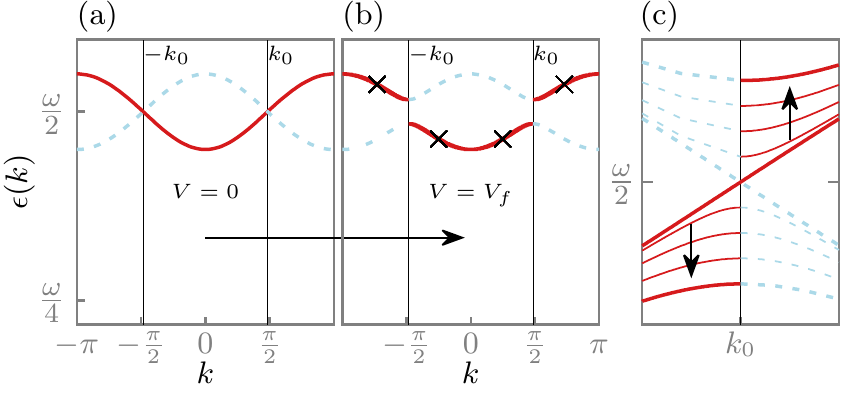}
		\caption{\label{fig:jump} Bulk Floquet spectrum as a function of momentum $k$ ($\mu=0.1 \pi$, $t=-\Delta=\pi$,  $\omega=\pi$, $\delta \mu_f=0.2$) prepared by adiabatically switching on the oscillating voltage $V$.  A band gap opens at energy $\omega/2$ (panel (a): undriven, panel (b): driven system). The quasi-energy of physical excitation, $\tilde{a}_k^\dagger\ket{\text{GS}}$, is marked as a red solid line, their `Bogoliubov shadow' as a blue dashed line. As physical excitations transform always into physical excitations under adiabatic evolution, the red solid line develops a jump at $\pm k_0$. This adiabatic evolution is shown in an enlarged view in panel (c).
			The resulting state is highly unstable as it allows the resonant creation of four quasi particles with total quasi-energy $\omega$. The 4 crosses give an example of such a process. 
		}
	\end{figure}

	Within the Floquet formalism, however, energies are only defined modulo $\omega$. Thus it is not possible 
	to distinguish physical excitations and their Bogoliubov shadows simply by their quasi-energy. 
	Fig.~\ref{fig:jump}a
	shows the spectrum of the undriven system folded into the first Floquet zone, $0\le E < \omega$. Here the red line is the physical excitation, the blue dashed line the Bogoliubov shadow $-E_{-k}+\omega$ projected into the Floquet zone $0<E\le \omega$. The modes touch at $k_0$ with  $E_{k_0}=-E_{-k_0}+\omega$, i.e., $E_{k_0}=\omega/2$.
	When the oscillating field is switched on, the modes hybridize, which is a necessary condition to create the $\pi$ Majorana mode at this quasi-energy, $\omega/2=\pi/T$. 
	
	Importantly, one has to track which of the hybridized modes describes a physical excitation and which describes only its Bogoliubov shadow.
	We can simply use adiabatic continuity to track whether an excitation is physical or a shadow excitation, see App.~\ref{app:bulkCalc}.
	This is shown in Fig.~\ref{fig:jump}b, where physical excitations are again shown in dark red while Bogoliubov shadows are depicted as dashed lines in light blue. For $k<k_0$ ($k>k_0$) the mode with quasi-energy smaller (larger) than $\omega/2$ has to be identified as a physical excitation shown in dark red. 
	At the crossing point $k_0$ a singularity develops leading to a jump in the nature of the excitation  (and the excitation energy), see Fig.~\ref{fig:jump}.
	An alternative, but equivalent, way to describe the same physics is to investigate the nature of the many-particle state created by the adiabatic evolution. The 	BCS wave function is written as $\prod_k (u_k + v_k a_k^\dagger a^\dagger_{-k}) |0\rangle$. Here the parameters $u_k$ and $v_k$ develop a singularity at $k=k_0$ where they exchange their role.

	To investigate the stability of the resulting state, we expand the interaction (last term in Eq.~\eqref{eq:Heff1})  in  creation and annihilation operators of the Floquet-Bogoliubov eigenmodes, $\tilde a^\dagger_\nu$ with quantum numbers $\nu$ (with $\nu=k$ for the translationally invariant bulk system considered above). 
	The interactions may trigger the creation of four quasi- particle  due to the term
	\begin{align}
		\Delta H_{\text{int}} =  \sum V^m_{\nu_1,\nu_2,\nu_3,\nu_4} e^{i m \omega \tau} \tilde a^\dagger_{\nu_1}\tilde a^\dagger_{\nu_2} \tilde a^\dagger_{\nu_3} \tilde a^\dagger_{\nu_4},
	\end{align} 
	where $V^m_{\nu_1,\nu_2,\nu_3,\nu_4}\sim \Delta^2/E_J$ is defined in App.~\ref{app:GoldenRule}. The corresponding creation rate of quasi particles can then be estimated using Fermi's golden rule
	\begin{align}
		\gamma_{qp}=4 \frac{2\pi}{\hbar}\sum_{\nu_i,m}\delta(E_{\nu_1}+E_{\nu_2}+E_{\nu_3}+E_{\nu_4}+ m \omega)|V^m_{\nu_1\nu_2\nu_3\nu_4}|^2, \label{eq:GoldenRule}
	\end{align} 
	where $E_{\nu_i}\ge 0$ are excitation energies obtained by diagonalizing the non-interacting Bogoliubov-Floquet system and the
	factor $4$ originates from the fact that four quasi-particles are created by the interaction. In our numerical calculations described below for finite-size systems
	with a discreet spectrum, we broaden the $\delta$-function  slightly by a box function of width $\Delta E$ to account for finite-lifetime effects, not captured by the golden-rule formula.
	
	Applying Eq.~\eqref{eq:GoldenRule} to the bulk energies, one realizes immediately that the jump in the quasi-energy of {\em physical} excitations enforces physical excitation both below and above $\frac{\omega}{2}$ (depicted in red in Fig.~\ref{fig:jump}). With the discontinuity, there is an abundance of resonant quasi-particle creation processes of the type depicted by four crosses in Fig.~\ref{fig:jump}, namely when quasi-energies add up to (multiples of) $\omega$. 
	In the bulk system, the total rate $\gamma_{qp}$ even diverges logarithmically due to processes occurring close to the location of the jump, $k_0$.
	We, therefore, conclude that the system is completely unstable when one prepares  a system with $\pi$ Majoranas simply by adiabatically switching on of the oscillating gate voltage. We expect that a similar statement applies to a wide range of models with $\pi$ Majoranas which naturally emerge from
	the crossing of Bogoliubov modes at energy $\omega/2$.

	\begin{figure}[t]
		\includegraphics[width=\columnwidth]{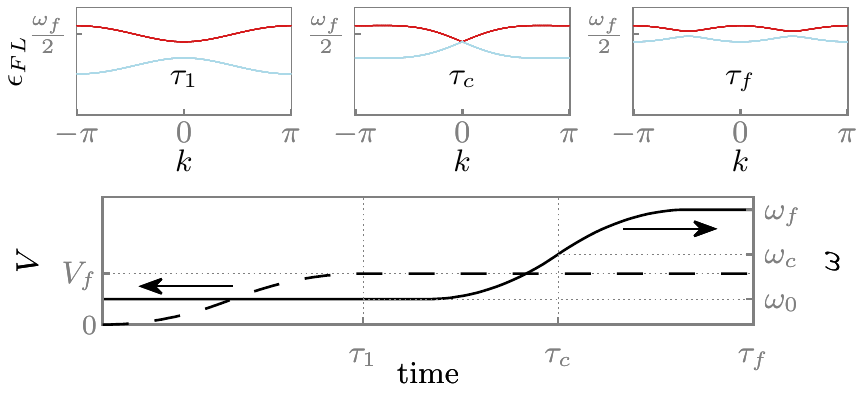}
		\caption{\label{fig:protocol} A `frequency sweep protocol' can be used to create a stable topological Floquet state.
			First, the amplitude of the oscillating voltage is increased slowly using a small frequency $\omega_0$. Then
			the frequency is increased up to $\omega_f$. At $\tau=\tau_c$ one enters the topological phase ($\mu=0.1\pi$, $t=-\pi$, $\Delta=\pi$, $V_f=0.2$, $\omega_0=0.8\pi$, $\omega_c=0.9\pi$, $\omega_f=\pi$).
			The Floquet spectra for $\tau=\tau_1$, $\tau_c$ and $\tau_f$ are shown on top. The final Floquet spectrum is identical to the one shown in Fig.~\ref{fig:jump}, but differs by the identification of physical excitations (red) vs. Bogoliubov shadows (blue).
		}
	\end{figure}

	Luckily, one can avoid this instability using a different preparation protocol, sketched in Fig.~\ref{fig:protocol}. The main idea is simply to avoid entering the topological Floquet phase directly by ramping up the oscillating field. Instead, we choose initially a frequency smaller than the band gap of the system while increasing the amplitude of the drive. Thus we avoid crossings of excitation energies and Bogoliubov shadows while preparing a non-topological Floquet state. Only at a second step, we slowly increase the frequency to reach the Floquet phase. 
	Level repulsion ensures that one never obtains a crossing of excitation- and shadow modes. Instead, they only touch at the quantum critical point (QCP), see 
	Fig.~\ref{fig:protocol}. Close to this point quasi-particle excitations are created by the Kibble-Zurek mechanism \cite{Polkovnikov2005,Zurek2005,Damski2005,Dziarmaga2005}.
Using the Landau-Zener formula for the Dirac Hamiltonian $m(t) \sigma_z+ v k \sigma_x$, we estimate the density $n_{\text{KZ}}$ of these quasi particles by   
$n_{\text{KZ}}=\frac{\sqrt{\partial_t m}}{v}$\cite{Landau1932,Zener1932}. For a system size of 10 in units of the correlation length $m/v$, one can easily reach a regime where in total less than one quasi particle is created for $\partial_t m < 0.01\,m^2$.

	\begin{figure}[!t]
		\includegraphics[width=\columnwidth]{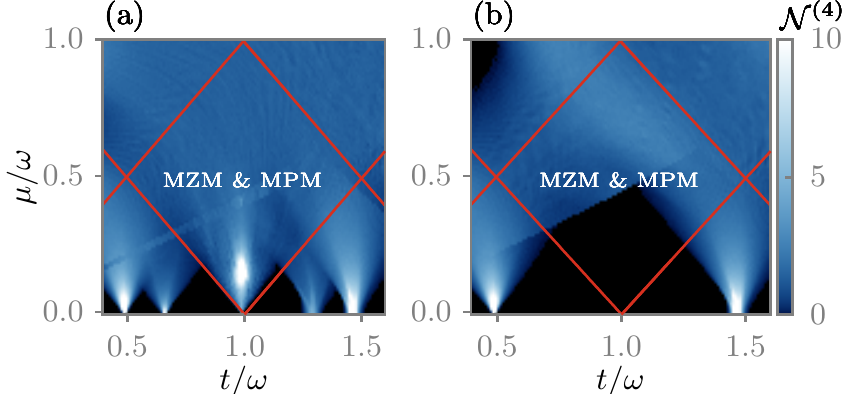}
		\caption{\label{fig:phaseNO}  Normalized phase space density $\mathcal N^{(4)}=\frac{1}{N^4}  \sum_{\nu_i,m}\delta(E_{\nu_1}+E_{\nu_2}+E_{\nu_3}+E_{\nu_4}+ m \omega)$ for the creation of four quasi particles out of the vacuum ($N=20$, $\delta E=0.02$, $t=-\Delta=\pi, \delta\mu=1$)
			Left: The Floquet state is prepared by adiabatically switching on the oscillating field. In this case the topological phase located inside the square is always unstable. Right: Preparation by the frequency-sweep protocol of Fig.~\ref{fig:protocol} (right). Here large parts of the topological phase remain stable as indicated by the black color.}
	\end{figure}

	With this protocol, the resulting Floquet phase is much more long-lived. In Fig.~\ref{fig:phaseNO} we show the phase space for the spontaneous creation of four quasi particles out of the vacuum for both types of protocols for a finite-size system. 
	The topological Floquet phase is {\em always} unstable if the state is prepared simply by an increase of the oscillating voltage. In contrast, there is a wide range of parameters where the topological phase is stable to leading order in the interaction if one uses the frequency-sweep protocol of Fig.~\ref{fig:protocol}. This analysis holds both for the long-ranged interactions derived in Eq.~\eqref{eq:Heff1} and for short-ranged interactions, see  App.~\ref{app:MatrixElements} for a numerical evaluation of the quasi-particle creation rate $\gamma_{qp}$, Eq.~\eqref{eq:GoldenRule}. We have checked that the region of stability is not enlarged when one considers only momentum-conserving bulk processes. As there is gap in the total 4-particle density of states, the   frequency-sweep protocol is furthermore stable against weak spatial disorder. Possible processes of 4th and higher order in the interactions, which involve the simultaneous creation of six or more quasi particles, may still exist but their prefactors will be strongly suppressed.
	
	\emph{Conclusion:} We have shown that by simply using an oscillating gate voltage, one can boost a Majorana box qubit into its Floquet version, thus increasing the number of logical qubits (in the fixed parity sector) from 1 to 3 combining Majorana zero modes and Majorana $\pi$ modes. Just increasing the oscillating field creates, however, a state which
	is completely unstable in the presence of interactions. To avoid this, one has to use a protocol that takes a detour: one first creates
	a non-topological Floquet state before entering the topological Floquet phase. This can, e.g., be realized by a frequency sweep.
	
	A Floquet Majorana box qubit provides with its three qubits a rich playground for quantum operations as outlined by Bomantara and Gong \cite{Bomantara2020}. Using a tunnel coupling of the edge modes via gated quantum dots one can realize a full set of Clifford gates, perform two-qubit operations with a help of an ancilla qubit, and test the braiding statistics of Majorana modes.
	 
	To summarize, our work shows that such Floquet devices can be operated despite the presence of interactions.

	\emph{Acknowledgements:}
	We acknowledge funding from the German Research
	Foundation (DFG) through CRC 183 (project number
	277101999, A01) and – under Germany’s Excellence Strategy – by the
	Cluster of Excellence Matter and Light for Quantum
	Computing (ML4Q) EXC2004/1 390534769. Furthermore, A.M. thanks
	the BCGS (Bonn-Cologne Graduate School of Physics
	and Astronomy) for support.
	E.B. acknowledges support from the Israel Science Foundation Quantum Science and Technology (grant no. 2074/19) and from a research grant from Irving and Cherna Moskowitz.
	
	\vspace{2cm}

	\bibliography{floquetMajoranaLiterature.bib}

\begin{thebibliography}{46}%
\makeatletter
\providecommand \@ifxundefined [1]{%
 \@ifx{#1\undefined}
}%
\providecommand \@ifnum [1]{%
 \ifnum #1\expandafter \@firstoftwo
 \else \expandafter \@secondoftwo
 \fi
}%
\providecommand \@ifx [1]{%
 \ifx #1\expandafter \@firstoftwo
 \else \expandafter \@secondoftwo
 \fi
}%
\providecommand \natexlab [1]{#1}%
\providecommand \enquote  [1]{``#1''}%
\providecommand \bibnamefont  [1]{#1}%
\providecommand \bibfnamefont [1]{#1}%
\providecommand \citenamefont [1]{#1}%
\providecommand \href@noop [0]{\@secondoftwo}%
\providecommand \href [0]{\begingroup \@sanitize@url \@href}%
\providecommand \@href[1]{\@@startlink{#1}\@@href}%
\providecommand \@@href[1]{\endgroup#1\@@endlink}%
\providecommand \@sanitize@url [0]{\catcode `\\12\catcode `\$12\catcode
  `\&12\catcode `\#12\catcode `\^12\catcode `\_12\catcode `\%12\relax}%
\providecommand \@@startlink[1]{}%
\providecommand \@@endlink[0]{}%
\providecommand \url  [0]{\begingroup\@sanitize@url \@url }%
\providecommand \@url [1]{\endgroup\@href {#1}{\urlprefix }}%
\providecommand \urlprefix  [0]{URL }%
\providecommand \Eprint [0]{\href }%
\providecommand \doibase [0]{https://doi.org/}%
\providecommand \selectlanguage [0]{\@gobble}%
\providecommand \bibinfo  [0]{\@secondoftwo}%
\providecommand \bibfield  [0]{\@secondoftwo}%
\providecommand \translation [1]{[#1]}%
\providecommand \BibitemOpen [0]{}%
\providecommand \bibitemStop [0]{}%
\providecommand \bibitemNoStop [0]{.\EOS\space}%
\providecommand \EOS [0]{\spacefactor3000\relax}%
\providecommand \BibitemShut  [1]{\csname bibitem#1\endcsname}%
\let\auto@bib@innerbib\@empty
\bibitem [{\citenamefont {Kitaev}(2001)}]{Kitaev2001}%
  \BibitemOpen
  \bibfield  {author} {\bibinfo {author} {\bibfnamefont {A.~Y.}\ \bibnamefont
  {Kitaev}},\ }\bibfield  {title} {\bibinfo {title} {{Unpaired Majorana
  fermions in quantum wires}},\ }\href
  {https://doi.org/10.1070/1063-7869/44/10S/S29} {\bibfield  {journal}
  {\bibinfo  {journal} {Phys. Usp.}\ }\textbf {\bibinfo {volume} {44}},\
  \bibinfo {pages} {131} (\bibinfo {year} {2001})},\ \Eprint
  {https://arxiv.org/abs/cond-mat/0010440} {arXiv:cond-mat/0010440}
  \BibitemShut {NoStop}%
\bibitem [{\citenamefont {Ivanov}(2001)}]{Ivanov2001}%
  \BibitemOpen
  \bibfield  {author} {\bibinfo {author} {\bibfnamefont {D.~A.}\ \bibnamefont
  {Ivanov}},\ }\bibfield  {title} {\bibinfo {title} {Non-abelian statistics of
  half-quantum vortices in $\mathit{p}$-wave superconductors},\ }\href
  {https://doi.org/10.1103/PhysRevLett.86.268} {\bibfield  {journal} {\bibinfo
  {journal} {Phys. Rev. Lett.}\ }\textbf {\bibinfo {volume} {86}},\ \bibinfo
  {pages} {268} (\bibinfo {year} {2001})}\BibitemShut {NoStop}%
\bibitem [{\citenamefont {Read}\ and\ \citenamefont {Green}(2000)}]{Read2000}%
  \BibitemOpen
  \bibfield  {author} {\bibinfo {author} {\bibfnamefont {N.}~\bibnamefont
  {Read}}\ and\ \bibinfo {author} {\bibfnamefont {D.}~\bibnamefont {Green}},\
  }\bibfield  {title} {\bibinfo {title} {Paired states of fermions in two
  dimensions with breaking of parity and time-reversal symmetries and the
  fractional quantum hall effect},\ }\href
  {https://doi.org/10.1103/PhysRevB.61.10267} {\bibfield  {journal} {\bibinfo
  {journal} {Phys. Rev. B}\ }\textbf {\bibinfo {volume} {61}},\ \bibinfo
  {pages} {10267} (\bibinfo {year} {2000})}\BibitemShut {NoStop}%
\bibitem [{\citenamefont {Fu}\ and\ \citenamefont {Kane}(2008)}]{Fu2008}%
  \BibitemOpen
  \bibfield  {author} {\bibinfo {author} {\bibfnamefont {L.}~\bibnamefont
  {Fu}}\ and\ \bibinfo {author} {\bibfnamefont {C.~L.}\ \bibnamefont {Kane}},\
  }\bibfield  {title} {\bibinfo {title} {{Superconducting Proximity Effect and
  Majorana Fermions at the Surface of a Topological Insulator}},\ }\href
  {https://doi.org/10.1103/physrevlett.100.096407} {\bibfield  {journal}
  {\bibinfo  {journal} {Physical Review Letters}\ }\textbf {\bibinfo {volume}
  {100}},\ \bibinfo {pages} {096407} (\bibinfo {year} {2008})}\BibitemShut
  {NoStop}%
\bibitem [{\citenamefont {Alicea}(2010)}]{Alicea2010}%
  \BibitemOpen
  \bibfield  {author} {\bibinfo {author} {\bibfnamefont {J.}~\bibnamefont
  {Alicea}},\ }\bibfield  {title} {\bibinfo {title} {{Majorana fermions in a
  tunable semiconductor device}},\ }\href
  {https://doi.org/10.1103/physrevb.81.125318} {\bibfield  {journal} {\bibinfo
  {journal} {Physical Review B}\ }\textbf {\bibinfo {volume} {81}},\ \bibinfo
  {pages} {125318} (\bibinfo {year} {2010})}\BibitemShut {NoStop}%
\bibitem [{\citenamefont {Oreg}\ \emph {et~al.}(2010)\citenamefont {Oreg},
  \citenamefont {Refael},\ and\ \citenamefont {von Oppen}}]{Oreg2010}%
  \BibitemOpen
  \bibfield  {author} {\bibinfo {author} {\bibfnamefont {Y.}~\bibnamefont
  {Oreg}}, \bibinfo {author} {\bibfnamefont {G.}~\bibnamefont {Refael}},\ and\
  \bibinfo {author} {\bibfnamefont {F.}~\bibnamefont {von Oppen}},\ }\bibfield
  {title} {\bibinfo {title} {{Helical Liquids and Majorana Bound States in
  Quantum Wires}},\ }\href {https://doi.org/10.1103/physrevlett.105.177002}
  {\bibfield  {journal} {\bibinfo  {journal} {Physical Review Letters}\
  }\textbf {\bibinfo {volume} {105}},\ \bibinfo {pages} {177002} (\bibinfo
  {year} {2010})}\BibitemShut {NoStop}%
\bibitem [{\citenamefont {Lutchyn}\ \emph {et~al.}(2010)\citenamefont
  {Lutchyn}, \citenamefont {Sau},\ and\ \citenamefont
  {Das~Sarma}}]{Lutchyn2010}%
  \BibitemOpen
  \bibfield  {author} {\bibinfo {author} {\bibfnamefont {R.~M.}\ \bibnamefont
  {Lutchyn}}, \bibinfo {author} {\bibfnamefont {J.~D.}\ \bibnamefont {Sau}},\
  and\ \bibinfo {author} {\bibfnamefont {S.}~\bibnamefont {Das~Sarma}},\
  }\bibfield  {title} {\bibinfo {title} {Majorana fermions and a topological
  phase transition in semiconductor-superconductor heterostructures},\ }\href
  {https://doi.org/10.1103/PhysRevLett.105.077001} {\bibfield  {journal}
  {\bibinfo  {journal} {Phys. Rev. Lett.}\ }\textbf {\bibinfo {volume} {105}},\
  \bibinfo {pages} {077001} (\bibinfo {year} {2010})}\BibitemShut {NoStop}%
\bibitem [{\citenamefont {Brouwer}\ \emph {et~al.}(2011)\citenamefont
  {Brouwer}, \citenamefont {Duckheim}, \citenamefont {Romito},\ and\
  \citenamefont {von Oppen}}]{Brouwer2011}%
  \BibitemOpen
  \bibfield  {author} {\bibinfo {author} {\bibfnamefont {P.~W.}\ \bibnamefont
  {Brouwer}}, \bibinfo {author} {\bibfnamefont {M.}~\bibnamefont {Duckheim}},
  \bibinfo {author} {\bibfnamefont {A.}~\bibnamefont {Romito}},\ and\ \bibinfo
  {author} {\bibfnamefont {F.}~\bibnamefont {von Oppen}},\ }\bibfield  {title}
  {\bibinfo {title} {{Topological superconducting phases in disordered quantum
  wires with strong spin-orbit coupling}},\ }\href
  {https://doi.org/10.1103/physrevb.84.144526} {\bibfield  {journal} {\bibinfo
  {journal} {Physical Review B}\ }\textbf {\bibinfo {volume} {84}},\ \bibinfo
  {pages} {144526} (\bibinfo {year} {2011})}\BibitemShut {NoStop}%
\bibitem [{\citenamefont {Manousakis}\ \emph {et~al.}(2017)\citenamefont
  {Manousakis}, \citenamefont {Altland}, \citenamefont {Bagrets}, \citenamefont
  {Egger},\ and\ \citenamefont {Ando}}]{Manousakis2017}%
  \BibitemOpen
  \bibfield  {author} {\bibinfo {author} {\bibfnamefont {J.}~\bibnamefont
  {Manousakis}}, \bibinfo {author} {\bibfnamefont {A.}~\bibnamefont {Altland}},
  \bibinfo {author} {\bibfnamefont {D.}~\bibnamefont {Bagrets}}, \bibinfo
  {author} {\bibfnamefont {R.}~\bibnamefont {Egger}},\ and\ \bibinfo {author}
  {\bibfnamefont {Y.}~\bibnamefont {Ando}},\ }\bibfield  {title} {\bibinfo
  {title} {{Majorana qubits in a topological insulator nanoribbon
  architecture}},\ }\href {https://doi.org/10.1103/physrevb.95.165424}
  {\bibfield  {journal} {\bibinfo  {journal} {Physical Review B}\ }\textbf
  {\bibinfo {volume} {95}},\ \bibinfo {pages} {165424} (\bibinfo {year}
  {2017})}\BibitemShut {NoStop}%
\bibitem [{\citenamefont {Lutchyn}\ \emph {et~al.}(2018)\citenamefont
  {Lutchyn}, \citenamefont {Bakkers}, \citenamefont {Kouwenhoven},
  \citenamefont {Krogstrup}, \citenamefont {Marcus},\ and\ \citenamefont
  {Oreg}}]{Lutchyn2018}%
  \BibitemOpen
  \bibfield  {author} {\bibinfo {author} {\bibfnamefont {R.~M.}\ \bibnamefont
  {Lutchyn}}, \bibinfo {author} {\bibfnamefont {E.~P. A.~M.}\ \bibnamefont
  {Bakkers}}, \bibinfo {author} {\bibfnamefont {L.~P.}\ \bibnamefont
  {Kouwenhoven}}, \bibinfo {author} {\bibfnamefont {P.}~\bibnamefont
  {Krogstrup}}, \bibinfo {author} {\bibfnamefont {C.~M.}\ \bibnamefont
  {Marcus}},\ and\ \bibinfo {author} {\bibfnamefont {Y.}~\bibnamefont {Oreg}},\
  }\bibfield  {title} {\bibinfo {title} {{Majorana zero modes in
  superconductor{\textendash}semiconductor heterostructures}},\ }\href
  {https://doi.org/10.1038/s41578-018-0003-1} {\bibfield  {journal} {\bibinfo
  {journal} {Nature Reviews Materials}\ }\textbf {\bibinfo {volume} {3}},\
  \bibinfo {pages} {52} (\bibinfo {year} {2018})}\BibitemShut {NoStop}%
\bibitem [{\citenamefont {Flensberg}\ \emph {et~al.}(2021)\citenamefont
  {Flensberg}, \citenamefont {von Oppen},\ and\ \citenamefont
  {Stern}}]{Flensberg2021}%
  \BibitemOpen
  \bibfield  {author} {\bibinfo {author} {\bibfnamefont {K.}~\bibnamefont
  {Flensberg}}, \bibinfo {author} {\bibfnamefont {F.}~\bibnamefont {von
  Oppen}},\ and\ \bibinfo {author} {\bibfnamefont {A.}~\bibnamefont {Stern}},\
  }\bibfield  {title} {\bibinfo {title} {{Engineered platforms for topological
  superconductivity and Majorana zero modes}},\ }\bibfield  {journal} {\bibinfo
   {journal} {Nature Reviews Materials}\ }\href
  {https://doi.org/10.1038/s41578-021-00336-6} {10.1038/s41578-021-00336-6}
  (\bibinfo {year} {2021})\BibitemShut {NoStop}%
\bibitem [{\citenamefont {Plugge}\ \emph {et~al.}(2017)\citenamefont {Plugge},
  \citenamefont {Rasmussen}, \citenamefont {Egger},\ and\ \citenamefont
  {Flensberg}}]{Plugge2017}%
  \BibitemOpen
  \bibfield  {author} {\bibinfo {author} {\bibfnamefont {S.}~\bibnamefont
  {Plugge}}, \bibinfo {author} {\bibfnamefont {A.}~\bibnamefont {Rasmussen}},
  \bibinfo {author} {\bibfnamefont {R.}~\bibnamefont {Egger}},\ and\ \bibinfo
  {author} {\bibfnamefont {K.}~\bibnamefont {Flensberg}},\ }\bibfield  {title}
  {\bibinfo {title} {{Majorana box qubits}},\ }\href
  {https://doi.org/10.1088/1367-2630/aa54e1} {\bibfield  {journal} {\bibinfo
  {journal} {New Journal of Physics}\ }\textbf {\bibinfo {volume} {19}},\
  \bibinfo {pages} {012001} (\bibinfo {year} {2017})}\BibitemShut {NoStop}%
\bibitem [{\citenamefont {Karzig}\ \emph {et~al.}(2017)\citenamefont {Karzig},
  \citenamefont {Knapp}, \citenamefont {Lutchyn}, \citenamefont {Bonderson},
  \citenamefont {Hastings}, \citenamefont {Nayak}, \citenamefont {Alicea},
  \citenamefont {Flensberg}, \citenamefont {Plugge}, \citenamefont {Oreg},
  \citenamefont {Marcus},\ and\ \citenamefont {Freedman}}]{Karzig2017}%
  \BibitemOpen
  \bibfield  {author} {\bibinfo {author} {\bibfnamefont {T.}~\bibnamefont
  {Karzig}}, \bibinfo {author} {\bibfnamefont {C.}~\bibnamefont {Knapp}},
  \bibinfo {author} {\bibfnamefont {R.~M.}\ \bibnamefont {Lutchyn}}, \bibinfo
  {author} {\bibfnamefont {P.}~\bibnamefont {Bonderson}}, \bibinfo {author}
  {\bibfnamefont {M.~B.}\ \bibnamefont {Hastings}}, \bibinfo {author}
  {\bibfnamefont {C.}~\bibnamefont {Nayak}}, \bibinfo {author} {\bibfnamefont
  {J.}~\bibnamefont {Alicea}}, \bibinfo {author} {\bibfnamefont
  {K.}~\bibnamefont {Flensberg}}, \bibinfo {author} {\bibfnamefont
  {S.}~\bibnamefont {Plugge}}, \bibinfo {author} {\bibfnamefont
  {Y.}~\bibnamefont {Oreg}}, \bibinfo {author} {\bibfnamefont {C.~M.}\
  \bibnamefont {Marcus}},\ and\ \bibinfo {author} {\bibfnamefont {M.~H.}\
  \bibnamefont {Freedman}},\ }\bibfield  {title} {\bibinfo {title} {Scalable
  designs for quasiparticle-poisoning-protected topological quantum computation
  with majorana zero modes},\ }\href
  {https://doi.org/10.1103/PhysRevB.95.235305} {\bibfield  {journal} {\bibinfo
  {journal} {Phys. Rev. B}\ }\textbf {\bibinfo {volume} {95}},\ \bibinfo
  {pages} {235305} (\bibinfo {year} {2017})}\BibitemShut {NoStop}%
\bibitem [{\citenamefont {Vijay}\ and\ \citenamefont {Fu}(2016)}]{Vijay2016}%
  \BibitemOpen
  \bibfield  {author} {\bibinfo {author} {\bibfnamefont {S.}~\bibnamefont
  {Vijay}}\ and\ \bibinfo {author} {\bibfnamefont {L.}~\bibnamefont {Fu}},\
  }\bibfield  {title} {\bibinfo {title} {{Teleportation-based quantum
  information processing with Majorana zero modes}},\ }\href
  {https://doi.org/10.1103/physrevb.94.235446} {\bibfield  {journal} {\bibinfo
  {journal} {Physical Review B}\ }\textbf {\bibinfo {volume} {94}},\ \bibinfo
  {pages} {235446} (\bibinfo {year} {2016})}\BibitemShut {NoStop}%
\bibitem [{\citenamefont {Kitagawa}\ \emph {et~al.}(2010)\citenamefont
  {Kitagawa}, \citenamefont {Berg}, \citenamefont {Rudner},\ and\ \citenamefont
  {Demler}}]{Kitagawa2010}%
  \BibitemOpen
  \bibfield  {author} {\bibinfo {author} {\bibfnamefont {T.}~\bibnamefont
  {Kitagawa}}, \bibinfo {author} {\bibfnamefont {E.}~\bibnamefont {Berg}},
  \bibinfo {author} {\bibfnamefont {M.}~\bibnamefont {Rudner}},\ and\ \bibinfo
  {author} {\bibfnamefont {E.}~\bibnamefont {Demler}},\ }\bibfield  {title}
  {\bibinfo {title} {Topological characterization of periodically driven
  quantum systems},\ }\href {https://doi.org/10.1103/PhysRevB.82.235114}
  {\bibfield  {journal} {\bibinfo  {journal} {Phys. Rev. B}\ }\textbf {\bibinfo
  {volume} {82}},\ \bibinfo {pages} {235114} (\bibinfo {year}
  {2010})}\BibitemShut {NoStop}%
\bibitem [{\citenamefont {Wilczek}(2012)}]{Wilczek2012}%
  \BibitemOpen
  \bibfield  {author} {\bibinfo {author} {\bibfnamefont {F.}~\bibnamefont
  {Wilczek}},\ }\bibfield  {title} {\bibinfo {title} {Quantum time crystals},\
  }\href {https://doi.org/10.1103/PhysRevLett.109.160401} {\bibfield  {journal}
  {\bibinfo  {journal} {Phys. Rev. Lett.}\ }\textbf {\bibinfo {volume} {109}},\
  \bibinfo {pages} {160401} (\bibinfo {year} {2012})}\BibitemShut {NoStop}%
\bibitem [{\citenamefont {Rudner}\ \emph {et~al.}(2013)\citenamefont {Rudner},
  \citenamefont {Lindner}, \citenamefont {Berg},\ and\ \citenamefont
  {Levin}}]{Rudner2013}%
  \BibitemOpen
  \bibfield  {author} {\bibinfo {author} {\bibfnamefont {M.~S.}\ \bibnamefont
  {Rudner}}, \bibinfo {author} {\bibfnamefont {N.~H.}\ \bibnamefont {Lindner}},
  \bibinfo {author} {\bibfnamefont {E.}~\bibnamefont {Berg}},\ and\ \bibinfo
  {author} {\bibfnamefont {M.}~\bibnamefont {Levin}},\ }\bibfield  {title}
  {\bibinfo {title} {Anomalous edge states and the bulk-edge correspondence for
  periodically driven two-dimensional systems},\ }\href
  {https://doi.org/10.1103/PhysRevX.3.031005} {\bibfield  {journal} {\bibinfo
  {journal} {Phys. Rev. X}\ }\textbf {\bibinfo {volume} {3}},\ \bibinfo {pages}
  {031005} (\bibinfo {year} {2013})}\BibitemShut {NoStop}%
\bibitem [{\citenamefont {Nathan}\ and\ \citenamefont
  {Rudner}(2015)}]{nathan2015topological}%
  \BibitemOpen
  \bibfield  {author} {\bibinfo {author} {\bibfnamefont {F.}~\bibnamefont
  {Nathan}}\ and\ \bibinfo {author} {\bibfnamefont {M.~S.}\ \bibnamefont
  {Rudner}},\ }\bibfield  {title} {\bibinfo {title} {Topological singularities
  and the general classification of floquet--bloch systems},\ }\href
  {https://iopscience.iop.org/article/10.1088/1367-2630/17/12/125014}
  {\bibfield  {journal} {\bibinfo  {journal} {New Journal of Physics}\ }\textbf
  {\bibinfo {volume} {17}},\ \bibinfo {pages} {125014} (\bibinfo {year}
  {2015})}\BibitemShut {NoStop}%
\bibitem [{\citenamefont {von Keyserlingk}\ and\ \citenamefont
  {Sondhi}(2016)}]{Keyserlingk2016}%
  \BibitemOpen
  \bibfield  {author} {\bibinfo {author} {\bibfnamefont {C.~W.}\ \bibnamefont
  {von Keyserlingk}}\ and\ \bibinfo {author} {\bibfnamefont {S.~L.}\
  \bibnamefont {Sondhi}},\ }\bibfield  {title} {\bibinfo {title} {Phase
  structure of one-dimensional interacting floquet systems. i. abelian
  symmetry-protected topological phases},\ }\href
  {https://doi.org/10.1103/PhysRevB.93.245145} {\bibfield  {journal} {\bibinfo
  {journal} {Phys. Rev. B}\ }\textbf {\bibinfo {volume} {93}},\ \bibinfo
  {pages} {245145} (\bibinfo {year} {2016})}\BibitemShut {NoStop}%
\bibitem [{\citenamefont {Else}\ and\ \citenamefont {Nayak}(2016)}]{Else2016}%
  \BibitemOpen
  \bibfield  {author} {\bibinfo {author} {\bibfnamefont {D.~V.}\ \bibnamefont
  {Else}}\ and\ \bibinfo {author} {\bibfnamefont {C.}~\bibnamefont {Nayak}},\
  }\bibfield  {title} {\bibinfo {title} {Classification of topological phases
  in periodically driven interacting systems},\ }\href
  {https://doi.org/10.1103/PhysRevB.93.201103} {\bibfield  {journal} {\bibinfo
  {journal} {Phys. Rev. B}\ }\textbf {\bibinfo {volume} {93}},\ \bibinfo
  {pages} {201103(R)} (\bibinfo {year} {2016})}\BibitemShut {NoStop}%
\bibitem [{\citenamefont {Else}\ \emph {et~al.}(2016)\citenamefont {Else},
  \citenamefont {Bauer},\ and\ \citenamefont {Nayak}}]{Else2016a}%
  \BibitemOpen
  \bibfield  {author} {\bibinfo {author} {\bibfnamefont {D.~V.}\ \bibnamefont
  {Else}}, \bibinfo {author} {\bibfnamefont {B.}~\bibnamefont {Bauer}},\ and\
  \bibinfo {author} {\bibfnamefont {C.}~\bibnamefont {Nayak}},\ }\bibfield
  {title} {\bibinfo {title} {Floquet time crystals},\ }\href
  {https://doi.org/10.1103/PhysRevLett.117.090402} {\bibfield  {journal}
  {\bibinfo  {journal} {Phys. Rev. Lett.}\ }\textbf {\bibinfo {volume} {117}},\
  \bibinfo {pages} {090402} (\bibinfo {year} {2016})}\BibitemShut {NoStop}%
\bibitem [{\citenamefont {Potter}\ \emph {et~al.}(2016)\citenamefont {Potter},
  \citenamefont {Morimoto},\ and\ \citenamefont {Vishwanath}}]{Potter2016}%
  \BibitemOpen
  \bibfield  {author} {\bibinfo {author} {\bibfnamefont {A.~C.}\ \bibnamefont
  {Potter}}, \bibinfo {author} {\bibfnamefont {T.}~\bibnamefont {Morimoto}},\
  and\ \bibinfo {author} {\bibfnamefont {A.}~\bibnamefont {Vishwanath}},\
  }\bibfield  {title} {\bibinfo {title} {Classification of interacting
  topological floquet phases in one dimension},\ }\href
  {https://doi.org/10.1103/PhysRevX.6.041001} {\bibfield  {journal} {\bibinfo
  {journal} {Phys. Rev. X}\ }\textbf {\bibinfo {volume} {6}},\ \bibinfo {pages}
  {041001} (\bibinfo {year} {2016})}\BibitemShut {NoStop}%
\bibitem [{\citenamefont {Titum}\ \emph {et~al.}(2016)\citenamefont {Titum},
  \citenamefont {Berg}, \citenamefont {Rudner}, \citenamefont {Refael},\ and\
  \citenamefont {Lindner}}]{Titum2016}%
  \BibitemOpen
  \bibfield  {author} {\bibinfo {author} {\bibfnamefont {P.}~\bibnamefont
  {Titum}}, \bibinfo {author} {\bibfnamefont {E.}~\bibnamefont {Berg}},
  \bibinfo {author} {\bibfnamefont {M.~S.}\ \bibnamefont {Rudner}}, \bibinfo
  {author} {\bibfnamefont {G.}~\bibnamefont {Refael}},\ and\ \bibinfo {author}
  {\bibfnamefont {N.~H.}\ \bibnamefont {Lindner}},\ }\bibfield  {title}
  {\bibinfo {title} {Anomalous floquet-anderson insulator as a nonadiabatic
  quantized charge pump},\ }\href {https://doi.org/10.1103/PhysRevX.6.021013}
  {\bibfield  {journal} {\bibinfo  {journal} {Phys. Rev. X}\ }\textbf {\bibinfo
  {volume} {6}},\ \bibinfo {pages} {021013} (\bibinfo {year}
  {2016})}\BibitemShut {NoStop}%
\bibitem [{\citenamefont {Roy}\ and\ \citenamefont {Harper}(2016)}]{Roy2016}%
  \BibitemOpen
  \bibfield  {author} {\bibinfo {author} {\bibfnamefont {R.}~\bibnamefont
  {Roy}}\ and\ \bibinfo {author} {\bibfnamefont {F.}~\bibnamefont {Harper}},\
  }\bibfield  {title} {\bibinfo {title} {Abelian floquet symmetry-protected
  topological phases in one dimension},\ }\href
  {https://doi.org/10.1103/PhysRevB.94.125105} {\bibfield  {journal} {\bibinfo
  {journal} {Phys. Rev. B}\ }\textbf {\bibinfo {volume} {94}},\ \bibinfo
  {pages} {125105} (\bibinfo {year} {2016})}\BibitemShut {NoStop}%
\bibitem [{\citenamefont {von Keyserlingk}\ \emph {et~al.}(2016)\citenamefont
  {von Keyserlingk}, \citenamefont {Khemani},\ and\ \citenamefont
  {Sondhi}}]{Khemani2016a}%
  \BibitemOpen
  \bibfield  {author} {\bibinfo {author} {\bibfnamefont {C.~W.}\ \bibnamefont
  {von Keyserlingk}}, \bibinfo {author} {\bibfnamefont {V.}~\bibnamefont
  {Khemani}},\ and\ \bibinfo {author} {\bibfnamefont {S.~L.}\ \bibnamefont
  {Sondhi}},\ }\bibfield  {title} {\bibinfo {title} {Absolute stability and
  spatiotemporal long-range order in floquet systems},\ }\href
  {https://doi.org/10.1103/PhysRevB.94.085112} {\bibfield  {journal} {\bibinfo
  {journal} {Phys. Rev. B}\ }\textbf {\bibinfo {volume} {94}},\ \bibinfo
  {pages} {085112} (\bibinfo {year} {2016})}\BibitemShut {NoStop}%
\bibitem [{\citenamefont {Jiang}\ \emph {et~al.}(2011)\citenamefont {Jiang},
  \citenamefont {Kitagawa}, \citenamefont {Alicea}, \citenamefont {Akhmerov},
  \citenamefont {Pekker}, \citenamefont {Refael}, \citenamefont {Cirac},
  \citenamefont {Demler}, \citenamefont {Lukin},\ and\ \citenamefont
  {Zoller}}]{Jiang2011}%
  \BibitemOpen
  \bibfield  {author} {\bibinfo {author} {\bibfnamefont {L.}~\bibnamefont
  {Jiang}}, \bibinfo {author} {\bibfnamefont {T.}~\bibnamefont {Kitagawa}},
  \bibinfo {author} {\bibfnamefont {J.}~\bibnamefont {Alicea}}, \bibinfo
  {author} {\bibfnamefont {A.~R.}\ \bibnamefont {Akhmerov}}, \bibinfo {author}
  {\bibfnamefont {D.}~\bibnamefont {Pekker}}, \bibinfo {author} {\bibfnamefont
  {G.}~\bibnamefont {Refael}}, \bibinfo {author} {\bibfnamefont {J.~I.}\
  \bibnamefont {Cirac}}, \bibinfo {author} {\bibfnamefont {E.}~\bibnamefont
  {Demler}}, \bibinfo {author} {\bibfnamefont {M.~D.}\ \bibnamefont {Lukin}},\
  and\ \bibinfo {author} {\bibfnamefont {P.}~\bibnamefont {Zoller}},\
  }\bibfield  {title} {\bibinfo {title} {Majorana fermions in equilibrium and
  in driven cold-atom quantum wires},\ }\href
  {https://doi.org/10.1103/PhysRevLett.106.220402} {\bibfield  {journal}
  {\bibinfo  {journal} {Phys. Rev. Lett.}\ }\textbf {\bibinfo {volume} {106}},\
  \bibinfo {pages} {220402} (\bibinfo {year} {2011})}\BibitemShut {NoStop}%
\bibitem [{\citenamefont {Liu}\ \emph {et~al.}(2012)\citenamefont {Liu},
  \citenamefont {Levchenko},\ and\ \citenamefont {Baranger}}]{Liu2012}%
  \BibitemOpen
  \bibfield  {author} {\bibinfo {author} {\bibfnamefont {D.~E.}\ \bibnamefont
  {Liu}}, \bibinfo {author} {\bibfnamefont {A.}~\bibnamefont {Levchenko}},\
  and\ \bibinfo {author} {\bibfnamefont {H.~U.}\ \bibnamefont {Baranger}},\
  }\bibfield  {title} {\bibinfo {title} {{Floquet Majorana Fermions for
  Topological Qubits}},\ }\href@noop {} {\bibfield  {journal} {\bibinfo
  {journal} {Phys. Rev. Lett. 111, 047002 (2013)}\ } (\bibinfo {year}
  {2012})},\ \Eprint {https://arxiv.org/abs/1211.1404} {arXiv:1211.1404
  [cond-mat.mes-hall]} \BibitemShut {NoStop}%
\bibitem [{\citenamefont {Kundu}\ and\ \citenamefont
  {Seradjeh}(2013)}]{Kundu2013}%
  \BibitemOpen
  \bibfield  {author} {\bibinfo {author} {\bibfnamefont {A.}~\bibnamefont
  {Kundu}}\ and\ \bibinfo {author} {\bibfnamefont {B.}~\bibnamefont
  {Seradjeh}},\ }\bibfield  {title} {\bibinfo {title} {Transport signatures of
  floquet majorana fermions in driven topological superconductors},\ }\href
  {https://doi.org/10.1103/PhysRevLett.111.136402} {\bibfield  {journal}
  {\bibinfo  {journal} {Phys. Rev. Lett.}\ }\textbf {\bibinfo {volume} {111}},\
  \bibinfo {pages} {136402} (\bibinfo {year} {2013})}\BibitemShut {NoStop}%
\bibitem [{\citenamefont {Li}\ \emph {et~al.}(2014)\citenamefont {Li},
  \citenamefont {Kundu}, \citenamefont {Zhong},\ and\ \citenamefont
  {Seradjeh}}]{Li2014}%
  \BibitemOpen
  \bibfield  {author} {\bibinfo {author} {\bibfnamefont {Y.}~\bibnamefont
  {Li}}, \bibinfo {author} {\bibfnamefont {A.}~\bibnamefont {Kundu}}, \bibinfo
  {author} {\bibfnamefont {F.}~\bibnamefont {Zhong}},\ and\ \bibinfo {author}
  {\bibfnamefont {B.}~\bibnamefont {Seradjeh}},\ }\bibfield  {title} {\bibinfo
  {title} {Tunable floquet majorana fermions in driven coupled quantum dots},\
  }\href {https://doi.org/10.1103/PhysRevB.90.121401} {\bibfield  {journal}
  {\bibinfo  {journal} {Phys. Rev. B}\ }\textbf {\bibinfo {volume} {90}},\
  \bibinfo {pages} {121401(R)} (\bibinfo {year} {2014})}\BibitemShut {NoStop}%
\bibitem [{\citenamefont {Liu}\ \emph {et~al.}(2019)\citenamefont {Liu},
  \citenamefont {Shabani},\ and\ \citenamefont {Mitra}}]{Liu2019}%
  \BibitemOpen
  \bibfield  {author} {\bibinfo {author} {\bibfnamefont {D.~T.}\ \bibnamefont
  {Liu}}, \bibinfo {author} {\bibfnamefont {J.}~\bibnamefont {Shabani}},\ and\
  \bibinfo {author} {\bibfnamefont {A.}~\bibnamefont {Mitra}},\ }\bibfield
  {title} {\bibinfo {title} {Floquet majorana zero and $\ensuremath{\pi}$ modes
  in planar josephson junctions},\ }\href
  {https://doi.org/10.1103/PhysRevB.99.094303} {\bibfield  {journal} {\bibinfo
  {journal} {Phys. Rev. B}\ }\textbf {\bibinfo {volume} {99}},\ \bibinfo
  {pages} {094303} (\bibinfo {year} {2019})}\BibitemShut {NoStop}%
\bibitem [{\citenamefont {Peng}\ \emph {et~al.}(2021)\citenamefont {Peng},
  \citenamefont {Haim}, \citenamefont {Karzig}, \citenamefont {Peng},\ and\
  \citenamefont {Refael}}]{Peng2021}%
  \BibitemOpen
  \bibfield  {author} {\bibinfo {author} {\bibfnamefont {C.}~\bibnamefont
  {Peng}}, \bibinfo {author} {\bibfnamefont {A.}~\bibnamefont {Haim}}, \bibinfo
  {author} {\bibfnamefont {T.}~\bibnamefont {Karzig}}, \bibinfo {author}
  {\bibfnamefont {Y.}~\bibnamefont {Peng}},\ and\ \bibinfo {author}
  {\bibfnamefont {G.}~\bibnamefont {Refael}},\ }\bibfield  {title} {\bibinfo
  {title} {Floquet majorana bound states in voltage-biased planar josephson
  junctions},\ }\href {https://doi.org/10.1103/PhysRevResearch.3.023108}
  {\bibfield  {journal} {\bibinfo  {journal} {Phys. Rev. Research}\ }\textbf
  {\bibinfo {volume} {3}},\ \bibinfo {pages} {023108} (\bibinfo {year}
  {2021})}\BibitemShut {NoStop}%
\bibitem [{\citenamefont {Peng}\ and\ \citenamefont {Refael}(2018)}]{Peng2018}%
  \BibitemOpen
  \bibfield  {author} {\bibinfo {author} {\bibfnamefont {Y.}~\bibnamefont
  {Peng}}\ and\ \bibinfo {author} {\bibfnamefont {G.}~\bibnamefont {Refael}},\
  }\bibfield  {title} {\bibinfo {title} {Time-quasiperiodic topological
  superconductors with majorana multiplexing},\ }\href
  {https://doi.org/10.1103/physrevb.98.220509} {\bibfield  {journal} {\bibinfo
  {journal} {Physical Review B}\ }\textbf {\bibinfo {volume} {98}},\ \bibinfo
  {pages} {220509(R)} (\bibinfo {year} {2018})}\BibitemShut {NoStop}%
\bibitem [{\citenamefont {Bauer}\ \emph {et~al.}(2019)\citenamefont {Bauer},
  \citenamefont {Pereg-Barnea}, \citenamefont {Karzig}, \citenamefont {Rieder},
  \citenamefont {Refael}, \citenamefont {Berg},\ and\ \citenamefont
  {Oreg}}]{Bauer2019}%
  \BibitemOpen
  \bibfield  {author} {\bibinfo {author} {\bibfnamefont {B.}~\bibnamefont
  {Bauer}}, \bibinfo {author} {\bibfnamefont {T.}~\bibnamefont {Pereg-Barnea}},
  \bibinfo {author} {\bibfnamefont {T.}~\bibnamefont {Karzig}}, \bibinfo
  {author} {\bibfnamefont {M.-T.}\ \bibnamefont {Rieder}}, \bibinfo {author}
  {\bibfnamefont {G.}~\bibnamefont {Refael}}, \bibinfo {author} {\bibfnamefont
  {E.}~\bibnamefont {Berg}},\ and\ \bibinfo {author} {\bibfnamefont
  {Y.}~\bibnamefont {Oreg}},\ }\bibfield  {title} {\bibinfo {title}
  {Topologically protected braiding in a single wire using floquet majorana
  modes},\ }\href {https://doi.org/10.1103/PhysRevB.100.041102} {\bibfield
  {journal} {\bibinfo  {journal} {Phys. Rev. B}\ }\textbf {\bibinfo {volume}
  {100}},\ \bibinfo {pages} {041102(R)} (\bibinfo {year} {2019})}\BibitemShut
  {NoStop}%
\bibitem [{\citenamefont {Bomantara}\ and\ \citenamefont
  {Gong}(2020)}]{Bomantara2020}%
  \BibitemOpen
  \bibfield  {author} {\bibinfo {author} {\bibfnamefont {R.~W.}\ \bibnamefont
  {Bomantara}}\ and\ \bibinfo {author} {\bibfnamefont {J.}~\bibnamefont
  {Gong}},\ }\bibfield  {title} {\bibinfo {title} {Measurement-only quantum
  computation with floquet majorana corner modes},\ }\href
  {https://doi.org/10.1103/PhysRevB.101.085401} {\bibfield  {journal} {\bibinfo
   {journal} {Phys. Rev. B}\ }\textbf {\bibinfo {volume} {101}},\ \bibinfo
  {pages} {085401} (\bibinfo {year} {2020})}\BibitemShut {NoStop}%
\bibitem [{\citenamefont {{Jens Koch and Terri M. Yu and Jay Gambetta and A. A.
  Houck and D. I. Schuster and J. Majer and Alexandre Blais and M. H. Devoret
  and S. M. Girvin and R. J. Schoelkopf}}(2007)}]{Koch2007}%
  \BibitemOpen
  \bibfield  {author} {\bibinfo {author} {\bibnamefont {{Jens Koch and Terri M.
  Yu and Jay Gambetta and A. A. Houck and D. I. Schuster and J. Majer and
  Alexandre Blais and M. H. Devoret and S. M. Girvin and R. J. Schoelkopf}}},\
  }\bibfield  {title} {\bibinfo {title} {Charge-insensitive qubit design
  derived from the cooper pair box},\ }\href
  {https://doi.org/10.1103/physreva.76.042319} {\bibfield  {journal} {\bibinfo
  {journal} {Physical Review A}\ }\textbf {\bibinfo {volume} {76}},\ \bibinfo
  {pages} {042319} (\bibinfo {year} {2007})}\BibitemShut {NoStop}%
\bibitem [{\citenamefont {Polkovnikov}(2005)}]{Polkovnikov2005}%
  \BibitemOpen
  \bibfield  {author} {\bibinfo {author} {\bibfnamefont {A.}~\bibnamefont
  {Polkovnikov}},\ }\bibfield  {title} {\bibinfo {title} {Universal adiabatic
  dynamics in the vicinity of a quantum critical point},\ }\href
  {https://doi.org/10.1103/physrevb.72.161201} {\bibfield  {journal} {\bibinfo
  {journal} {Physical Review B}\ }\textbf {\bibinfo {volume} {72}},\ \bibinfo
  {pages} {161201(R)} (\bibinfo {year} {2005})}\BibitemShut {NoStop}%
\bibitem [{\citenamefont {Zurek}\ \emph {et~al.}(2005)\citenamefont {Zurek},
  \citenamefont {Dorner},\ and\ \citenamefont {Zoller}}]{Zurek2005}%
  \BibitemOpen
  \bibfield  {author} {\bibinfo {author} {\bibfnamefont {W.~H.}\ \bibnamefont
  {Zurek}}, \bibinfo {author} {\bibfnamefont {U.}~\bibnamefont {Dorner}},\ and\
  \bibinfo {author} {\bibfnamefont {P.}~\bibnamefont {Zoller}},\ }\bibfield
  {title} {\bibinfo {title} {Dynamics of a quantum phase transition},\ }\href
  {https://doi.org/10.1103/physrevlett.95.105701} {\bibfield  {journal}
  {\bibinfo  {journal} {Physical Review Letters}\ }\textbf {\bibinfo {volume}
  {95}},\ \bibinfo {pages} {105701} (\bibinfo {year} {2005})}\BibitemShut
  {NoStop}%
\bibitem [{\citenamefont {Damski}(2005)}]{Damski2005}%
  \BibitemOpen
  \bibfield  {author} {\bibinfo {author} {\bibfnamefont {B.}~\bibnamefont
  {Damski}},\ }\bibfield  {title} {\bibinfo {title} {The simplest quantum model
  supporting the kibble-zurek mechanism of topological defect production:
  Landau-zener transitions from a new perspective},\ }\href
  {https://doi.org/10.1103/physrevlett.95.035701} {\bibfield  {journal}
  {\bibinfo  {journal} {Physical Review Letters}\ }\textbf {\bibinfo {volume}
  {95}},\ \bibinfo {pages} {035701} (\bibinfo {year} {2005})}\BibitemShut
  {NoStop}%
\bibitem [{\citenamefont {Dziarmaga}(2005)}]{Dziarmaga2005}%
  \BibitemOpen
  \bibfield  {author} {\bibinfo {author} {\bibfnamefont {J.}~\bibnamefont
  {Dziarmaga}},\ }\bibfield  {title} {\bibinfo {title} {Dynamics of a quantum
  phase transition: Exact solution of the quantum ising model},\ }\href
  {https://doi.org/10.1103/physrevlett.95.245701} {\bibfield  {journal}
  {\bibinfo  {journal} {Physical Review Letters}\ }\textbf {\bibinfo {volume}
  {95}},\ \bibinfo {pages} {245701} (\bibinfo {year} {2005})}\BibitemShut
  {NoStop}%
\bibitem [{\citenamefont {Landau}(1932)}]{Landau1932}%
  \BibitemOpen
  \bibfield  {author} {\bibinfo {author} {\bibfnamefont {L.}~\bibnamefont
  {Landau}},\ }\bibfield  {title} {\bibinfo {title} {On the theory of transfer
  of energy at collisions ii},\ }\href@noop {} {\bibfield  {journal} {\bibinfo
  {journal} {Phys. Z. Sowjetunion}\ }\textbf {\bibinfo {volume} {2}},\ \bibinfo
  {pages} {118} (\bibinfo {year} {1932})}\BibitemShut {NoStop}%
\bibitem [{\citenamefont {Zener}(1932)}]{Zener1932}%
  \BibitemOpen
  \bibfield  {author} {\bibinfo {author} {\bibfnamefont {C.}~\bibnamefont
  {Zener}},\ }\bibfield  {title} {\bibinfo {title} {Non-adiabatic crossing of
  energy levels},\ }\href {https://doi.org/10.1098/rspa.1932.0165} {\bibfield
  {journal} {\bibinfo  {journal} {Proceedings of the Royal Society of London.
  Series A, Containing Papers of a Mathematical and Physical Character}\
  }\textbf {\bibinfo {volume} {137}},\ \bibinfo {pages} {696} (\bibinfo {year}
  {1932})}\BibitemShut {NoStop}%
\bibitem [{\citenamefont {Seetharam}\ \emph {et~al.}(2015)\citenamefont
  {Seetharam}, \citenamefont {Bardyn}, \citenamefont {Lindner}, \citenamefont
  {Rudner},\ and\ \citenamefont {Refael}}]{Seetharam2015}%
  \BibitemOpen
  \bibfield  {author} {\bibinfo {author} {\bibfnamefont {K.~I.}\ \bibnamefont
  {Seetharam}}, \bibinfo {author} {\bibfnamefont {C.-E.}\ \bibnamefont
  {Bardyn}}, \bibinfo {author} {\bibfnamefont {N.~H.}\ \bibnamefont {Lindner}},
  \bibinfo {author} {\bibfnamefont {M.~S.}\ \bibnamefont {Rudner}},\ and\
  \bibinfo {author} {\bibfnamefont {G.}~\bibnamefont {Refael}},\ }\bibfield
  {title} {\bibinfo {title} {{Controlled Population of Floquet-Bloch States via
  Coupling to Bose and Fermi Baths}},\ }\href
  {https://doi.org/10.1103/PhysRevX.5.041050} {\bibfield  {journal} {\bibinfo
  {journal} {Phys. Rev. X}\ }\textbf {\bibinfo {volume} {5}},\ \bibinfo {pages}
  {041050} (\bibinfo {year} {2015})}\BibitemShut {NoStop}%
\bibitem [{\citenamefont {Bilitewski}\ and\ \citenamefont
  {Cooper}(2015)}]{Bilitewski2015}%
  \BibitemOpen
  \bibfield  {author} {\bibinfo {author} {\bibfnamefont {T.}~\bibnamefont
  {Bilitewski}}\ and\ \bibinfo {author} {\bibfnamefont {N.~R.}\ \bibnamefont
  {Cooper}},\ }\bibfield  {title} {\bibinfo {title} {{Scattering theory for
  Floquet-Bloch states}},\ }\href {https://doi.org/10.1103/PhysRevA.91.033601}
  {\bibfield  {journal} {\bibinfo  {journal} {Phys. Rev. A}\ }\textbf {\bibinfo
  {volume} {91}},\ \bibinfo {pages} {033601} (\bibinfo {year}
  {2015})}\BibitemShut {NoStop}%
\bibitem [{\citenamefont {Genske}\ and\ \citenamefont
  {Rosch}(2015)}]{Genske2015}%
  \BibitemOpen
  \bibfield  {author} {\bibinfo {author} {\bibfnamefont {M.}~\bibnamefont
  {Genske}}\ and\ \bibinfo {author} {\bibfnamefont {A.}~\bibnamefont {Rosch}},\
  }\bibfield  {title} {\bibinfo {title} {{Floquet-Boltzmann equation for
  periodically driven Fermi systems}},\ }\href
  {https://doi.org/10.1103/PhysRevA.92.062108} {\bibfield  {journal} {\bibinfo
  {journal} {Phys. Rev. A}\ }\textbf {\bibinfo {volume} {92}},\ \bibinfo
  {pages} {062108} (\bibinfo {year} {2015})}\BibitemShut {NoStop}%
\bibitem [{\citenamefont {Rudner}\ and\ \citenamefont
  {Lindner}(2020)}]{Rudner2020}%
  \BibitemOpen
  \bibfield  {author} {\bibinfo {author} {\bibfnamefont {M.~S.}\ \bibnamefont
  {Rudner}}\ and\ \bibinfo {author} {\bibfnamefont {N.~H.}\ \bibnamefont
  {Lindner}},\ }\bibfield  {title} {\bibinfo {title} {{The Floquet Engineer's
  Handbook}},\ }\href@noop {} {\bibfield  {journal} {\bibinfo  {journal} {arXiv
  preprint arXiv:2003.08252}\ } (\bibinfo {year} {2020})},\ \Eprint
  {https://arxiv.org/abs/2003.08252} {arXiv:2003.08252 [cond-mat.mes-hall]}
  \BibitemShut {NoStop}%
\bibitem [{\citenamefont {Zirnbauer}(2021)}]{Zirnbauer2021}%
  \BibitemOpen
  \bibfield  {author} {\bibinfo {author} {\bibfnamefont {M.~R.}\ \bibnamefont
  {Zirnbauer}},\ }\bibfield  {title} {\bibinfo {title} {{Particle–hole
  symmetries in condensed matter}},\ }\href {https://doi.org/10.1063/5.0035358}
  {\bibfield  {journal} {\bibinfo  {journal} {Journal of Mathematical Physics}\
  }\textbf {\bibinfo {volume} {62}},\ \bibinfo {pages} {021101} (\bibinfo
  {year} {2021})},\ \Eprint
  {https://arxiv.org/abs/https://doi.org/10.1063/5.0035358}
  {https://doi.org/10.1063/5.0035358} \BibitemShut {NoStop}%
\end{thebibliography}%

	\begin{widetext}
		\appendix
		
		\section{Effective low-energy Hamiltonian}
		
		\label{app:A}\label{app:SWtrafo}

		In the following, we show how an effective low-energy Hamiltonian can be obtained starting from Eq.~\eqref{eq:H} of the main text.		
		We first apply the harmonic oscillator approximation $-E_J \cos \hat \phi \approx E_J (-1+\frac{1}{2}{\hat \phi}^2)  $ in the limit $E_J \gg E_c$ such that $\hat{ \phi}$ is pinned to one of the minima of $\cos(\hat{ \phi})$ and weakly fluctuates around the minimum.   
		We then apply a series of unitary transformations to the Hamiltonian 
		\begin{align}
			H_0(\tau)&=\frac{E_c}{2} \hat{Q}^2 + \frac{E_J}{2} {\hat{ \phi}}^2+\hat{Q} V(\tau)+\frac{\Delta}{2}\left(\sum_i c^\dagger_i c^\dagger_{i+1} e^{-i\hat{\phi}/2} +h.c.\right),\label{eq:H0app}
		\end{align}
		where we omitted  in Eq.~\eqref{eq:H0app} all terms from Eq. (1) of the main text not affected by the transformations considered below. Furthermore, we consider only one of the chains (the result can easily be generalized to the two chain case).
		
		To eliminate the time-dependent term linear in $\hat Q$, we apply the transformation
		$
		U_1(\tau)=e^{-i\hat{\phi}V(\tau)/(2E_c)}
		$ and arrive at
		\begin{align}
			H_1(\tau)&=U^\dagger_1(\tau)H_0(\tau)U_1(\tau)-i U^\dagger_1(\tau) \partial_\tau U_1(\tau) \nonumber \\
			&=\frac{E_c}{2} \hat{Q}^2 + \frac{E_J}{2}\hat{\phi}^2 +\frac{\Delta}{2}\left(\sum_i c^\dagger_i c^\dagger_{i+1} e^{-i\hat{\phi}/2} +h.c.\right) -\hat{\phi}\frac{\dot{V}(\tau)}{2 E_c}+\text{const.}.
		\end{align}
		Now we eliminate the offset in $\hat{\phi}$ with a second transformation
		$	U_2(\tau)=e^{-i\hat{Q}\frac{\dot{V}(\tau)}{4 E_c E_J}}
		$ and obtain
		\begin{align}
			H_2(\tau)&=U^\dagger_2(\tau)H_1(\tau)U_2(\tau)-i U_2^\dagger(\tau) \partial_\tau U_2(\tau)\nonumber \\
			&=\frac{E_c}{2} \hat{Q}^2 + \frac{E_J}{2}\hat{\phi}^2 +\frac{\Delta}{2}\left(\sum_i c^\dagger_i c^\dagger_{i+1} e^{-i(\hat{\phi}/2+\frac{\dot{V}(\tau)}{2E_c E_J})} +h.c.\right)-\hat{Q} \frac{V''(\tau)}{4 E_c E_J} +\text{const.}.
		\end{align}
		Hereafter, we simply neglect the tiny term proportional to $\frac{V''(\tau)}{ E_c E_J}\sim \frac{\omega^2 V_0}{E_c E_J} \ll \omega \ll E_c,E_J$.
		Next we expand $e^{-i \hat{\phi}/2}\approx 1-i \frac{\hat{ \phi}}{2}$ and obtain
		\begin{align}
			H_3(\tau)&=
			\frac{E_c}{2} \hat{Q}^2 + \frac{E_J}{2}\hat{\phi}^2 +\frac{\Delta}{2}\left(\sum_i c^\dagger_i c^\dagger_{i+1} e^{-i\frac{\dot{V(\tau)}}{2E_c E_J}} +h.c.\right)- \frac{\Delta}{4} T \hat{ \phi}+\text{const.} 
		\end{align}
		with \begin{equation}
			T=i\left(e^{-i\frac{\dot{V}(\tau)}{2 E_c E_J}}\sum_i c^\dagger_i c^\dagger_{i+1}- e^{i\frac{\dot{V}(\tau)}{2E_c E_J}}\sum_i c_{i+1} c_{i}\right).
		\end{equation}
		The displacement of $\hat{\phi}$ is eliminated by a last transformation  $U_3=e^{-i\hat{Q}\frac{\Delta}{8E_J} T }$. Since $U_3$ does not commute with the Kitaev Hamiltonian $H_k=\mu \sum_i c^\dagger_i c_i- \sum_i \left( \frac{t}{2}  c^\dagger_i c_{i+1}+ \frac{\Delta}{2} c^\dagger_i c^\dagger_{i+1} e^{-i\frac{\dot{V}(\tau)}{2E_c E_J}} +h.c.\right)$, we again include all terms in the Hamiltonian $\tilde{H}_3$

		\begin{align}
			H_4(\tau)&= U^\dagger_3(\tau)\tilde{H}_3(\tau)U_3(\tau)-i U_3^\dagger(\tau) \partial_\tau U_3(\tau)\nonumber \\
			&= \frac{E_c}{2} \hat{Q}^2 + \frac{E_J}{2}\hat{\phi}^2
			+H_k+\frac{\Delta^2}{32E_J}T^2+ \frac{\Delta}{8E_J}[T,H_k] \hat{Q}+\mathcal{O}(\frac{1}{E_J^2})+\text{const.}.
		\end{align}

		We first neglect the terms of order $\frac{1}{E_J^2}$.  The fifth term correspond to a small shift in $\hat{Q}$ which is of order $\sim \frac{\Delta^2}{E_J E_c} \ll E_c$, and thus can be neglected as well. Finally, we arrive at the low energy Hamiltonian $H_{\rm eff}$ (Eq.~\ref{eq:Heff1} in the main text)
		
		\begin{align}
			H_{\rm eff}=\mu \sum_i c^\dagger_i c_i- \frac{t}{2} \sum_i c^\dagger_i c_{i+1} +\frac{\Delta}{2}  e^{-i\frac{\dot{V}(\tau)}{2 E_c E_J}}\sum_i c^\dagger_i c^\dagger_{i+1} +h.c.+\frac{\Delta^2}{32E_J}T^2.
		\end{align} 
		\section{Floquet Fermi Golden Rule} 
		\label{app:GoldenRule}
		To study the interaction effects, we look at the Floquet Fermi Golden rule \cite{Seetharam2015,Bilitewski2015,Genske2015}. First the creation (annihilation) operators $c^\dagger_i$ ($c_i$) are expressed in terms of the Floquet operators $\tilde{a}^\dagger_\nu$ ($\tilde{a}_\nu$) where $\tilde{a}^\dagger_{\nu} (\tau)\ket{0}=\ket{\phi_{\nu}(\tau)}$ creates a fermion in a Floquet eigenstate. Here $\ket{0}$ is the quasi-particle vacuum. We use that the single-particle Floquet eigenstates of our Hamiltonian $\ket{\phi_\nu(\tau)}$ at equal times form a complete set in the physical Hilbert space $\mathbb{1}=\sum_\nu \ket{\phi_{\nu}(\tau)}\bra{\phi_{\nu}(\tau)}$, where the sum runs only over states of the first Floquet zone.
		Therefore,
		\begin{equation}
			c_j^\dagger\ket{0}=\sum_\nu \ket{\phi_{\nu}(\tau)}\bra{\phi_{\nu}(\tau)}\ket{j}=\sum_\nu \bra{\phi_{\nu}(\tau)}\ket{j} \tilde{a}^\dagger_\nu(\tau) \ket{0}
		\end{equation}
		and $c^\dagger_j=\sum_\nu \bar{\phi}_\nu(t,j)  \tilde{a}^\dagger_\nu(\tau)$ with $\bar{\phi}_\nu(t,j)=\bra{\phi_{\nu}(\tau)}\ket{j}$.
		Now we can use the mode expansion of Floquet states $\ket{\phi_\nu(\tau)}=\sum_n e^{-in\omega \tau }\ket{\phi_{\nu}^n}$ and see that the Floquet operators have an explicit time dependence in the Schr\"odinger picture. However, we are looking for an expression for the creation (annihilation) operators in terms of the Floquet operators. Therefore, let us expand the $\bar{\phi}_\nu(t,i)$ and we get with $\bar{\phi}_\nu^n(j)=\bra{\phi^n_\nu}\ket{j}$:
		\begin{equation}
			c^\dagger_j=\sum_\nu e^{in\omega \tau}\bar{\phi}_\nu^n(j)  \tilde{a}^\dagger_\nu(\tau)
		\end{equation}
		with $\tilde{a}^\dagger_\nu(\tau)=e^{i\epsilon_\nu \tau}\tilde{a}^\dagger_\nu$.
		Now we can write
		\begin{align}
			A(i\rightarrow f,\tau)&=\bra{\psi_f(\tau)}U(0,\tau)\ket{\psi_i(\tau=0)}\\
			&=-\frac{i}{\hbar}\int_0^\tau d\tau'\bra{\psi_f(0)} U(\tau',0)
			\Delta H_{\text{int}}U(0,\tau')\ket{\psi_i(0)} \nonumber\\
			&=-\frac{i \Delta^2}{32 \hbar E_J}\sum_{j,k}\sum_{n_1,n_2,n_3,n_4}\sum_{\nu_1,\nu_2,\nu_3,\nu_4} \int_0^\tau d\tau'  e^{i(n_1+n_2+n_3+n_4)\omega \tau'}e^{i(\epsilon_{\nu_1}+\epsilon_{\nu_2}+\epsilon_{\nu_3}+\epsilon_{\nu_4}) \tau'} 
			\nonumber \\
			&\times \bar{\phi}_{\nu_1}^{n_1}(j)\bar{\phi}_{\nu_2}^{n_2}(j+1)\bar{\phi}_{\nu_3}^{n_3}(k)\bar{\phi}_{\nu_4}^{n_4}(k+1) \bra{\psi_f(0)} \tilde{a}^\dagger_{\nu_1}\tilde{a}^\dagger_{\nu_2} \tilde{a}^\dagger_{\nu_3}\tilde{a}^\dagger_{\nu_4} \ket{\psi_i(0)} \nonumber\\
			&=\frac{\Delta^2}{32 E_J}\sum_{j,k}\sum_{n_1,n_2,n_3,n_4}\sum_{\nu_1,\nu_2,\nu_3,\nu_4}\frac{e^{i((n_1+n_2+n_3+n_4)\omega +\epsilon_{\nu_1}+\epsilon_{\nu_2}+\epsilon_{\nu_3}+\epsilon_{\nu_4}) \tau}-1}{(n_1+n_2+n_3+n_4)\omega +\epsilon_{\nu_1}+\epsilon_{\nu_2}+\epsilon_{\nu_3}+\epsilon_{\nu_4}} \nonumber\\
			&\times \bar{\phi}_{\nu_1}^{n_1}(j)\bar{\phi}_{\nu_2}^{n_2}(j+1)\bar{\phi}_{\nu_3}^{n_3}(k)\bar{\phi}_{\nu_4}^{n_4}(k+1)\bra{\psi_f(0)} \tilde{a}^\dagger_{\nu_1}\tilde{a}^\dagger_{\nu_2} \tilde{a}^\dagger_{\nu_3}\tilde{a}^\dagger_{\nu_4} \ket{\psi_i(0)}, \nonumber
		\end{align}
		where $U$ is the time-evolution operator and $\Delta H_{\text{int}}=\frac{\Delta^2}{32E_J}T^2$ describes the interaction meditated by the bulk superconductor. Starting from the adiabatic groundstate, only terms contribute where four quasi particles are created.
		With $\gamma_{i\rightarrow f}=\lim_{\tau \rightarrow\infty}\frac{P(i\rightarrow f,\tau)}{\tau}$ and the transition probability $P(i\rightarrow f,\tau )=|A(i\rightarrow f,\tau )|^2$, we obtain
		\begin{align}
			\gamma_{i\rightarrow f}
			&=\frac{2\pi}{\hbar}\sum_{\nu_1,\nu_2,\nu_3,\nu_4,n}\delta(\epsilon_{\nu_1}+\epsilon_{\nu_2}+\epsilon_{\nu_3}+\epsilon_{\nu_4}+ n \omega)|V^n_{\nu_1\nu_2\nu_3\nu_4}|^2
		\end{align}
		with
		$
		V^n_{\nu_1\nu_2\nu_3\nu_4}=\frac{\Delta^2}{32 E_J} \sum_{j,k}\sum_{n_1,n_2,n_3,n_4} \delta(n-(n_1+n_2+n_3+n_4))\bar{\phi}_{\nu_1}^{n_1}(j)\bar{\phi}_{\nu_2}^{n_2}(j+1)\bar{\phi}_{\nu_3}^{n_3}(k)\bar{\phi}_{\nu_4}^{n_4}(k+1)
		$. As four quasi particles are created in the process described above, the quasi-particle creation rate $\gamma_{qp}$, Eq.~\ref{eq:GoldenRule}, is given by $\gamma_{qp}=4\gamma_{i\rightarrow f}$.
		
		\section{Numerical calculation of the quasi-particle creation rates}\label{app:MatrixElements}
		
		In this section, we use Eq.~\eqref{eq:GoldenRule}  to calculate $\gamma_{qp}$ numerically, which describes the rate at which quasi particles are created in a finite size system. Our calculation thus includes both bulk and boundary processes.
		We replace the $\delta$ function in Eq.~\eqref{eq:GoldenRule} by a box with width $\delta E$ of the order of the single-particle level spacing to account for spectral broadening. The Floquet states and the resulting matrix elements are calculated by numerical diagonalization  typically taking into account 7 Floquet modes, which turns out to be sufficient  for convergence of the eigenenergies in the investigated parameter ranges. In Fig. \ref{fig:appscatRate} we show $\gamma_{qp}$ for a protocol where the initial state is either prepared by adiabatically switching on the oscillating gate voltage (red triangles) or by  the 
		frequency-sweep protocol discussed in the main text (black circles). In the grey-shaded regime, where both 
		Majorana zero modes and Majorana $\pi$ modes exist, the adiabatic protocol prepares a highly instable state with a large quasi-particle creation rate. In contrast, the system remains stable within the frequency-sweep protocol. In this case, $\gamma_{qp}$ calculated within the Golden-Rule approximation
		vanishes exactly in the topological phase for the chosen parameters. Note that higher-order processes involving the simultaneous creation of 6 or 8 quasi particles may still occur with a strongly suppressed prefactor.

		\begin{figure}
			\includegraphics[width=0.5\columnwidth]{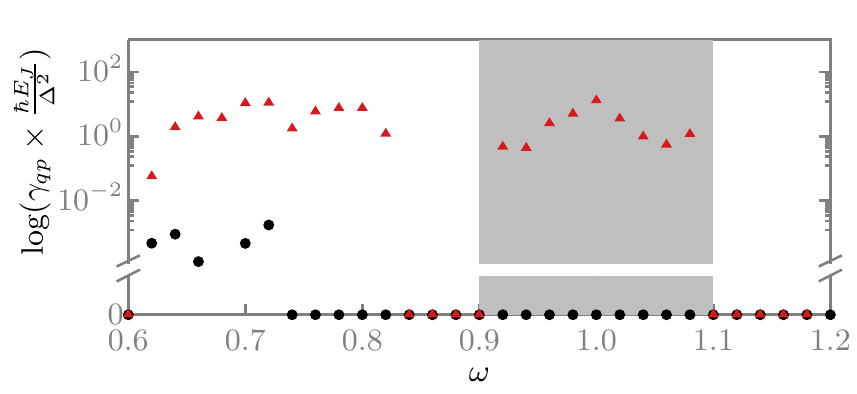}
			\caption{\label{fig:appscatRate}Quasi-particle creation rate $\gamma_{qp}$, computed for $N=20$ by replacing the $\delta$ function by a box of width $\delta E=0.02$, $t=-\Delta=\pi, \delta\mu=1, m_{max}=3$) for creating four quasi particles out of the vacuum using 
				an adiabatically prepared Floquet system (red) and a system prepared by the frequency-sweep protocol of Fig.~\ref{fig:protocol} (black). The grey shaded area shows the range where both Majorana zero modes and Majorana $\pi$ modes are available.}
		\end{figure}

		\section{Physical excitations and adiabatic continuity}\label{app:bulkCalc}
		In this section, we calculate analytically the Floquet-Bogoliubov spectrum in the bulk for small amplitudes of the oscillating potential.
		Our starting point is the {\em static}  BCS Hamiltonian describing a single wire in the absence of the oscillating gate voltage.
		\begin{equation}
			H_{\text{BdG}}=\frac{1}{2} \sum_{k}( c^\dagger_k, c_{-k})\left(
			\begin{array}{cc}
				-(\mu+t\cos k)  & -i \Delta \sin k \\
				i \Delta \sin k & \mu+t\cos k \\
			\end{array}
			\right) \left( \begin{array}{c}
				c_k\\
				c^\dagger_{-k}
			\end{array}\right)
		\end{equation}
		Using a Bogoliubov transformation, we diagonalize the Hamiltonian
		\begin{equation}
			H_{BdG}=\frac{1}{2} \sum_k(a^\dagger_{k}, a_{-k})\left(
			\begin{array}{cc}
				E_k& 0 \\
				0 & -E_k \\
			\end{array}
			\right) \left( \begin{array}{c}
				a_{k}\\
				a^\dagger_{-k}
			\end{array}\right) = \sum_k E_k a^\dagger_{k} a_{k}
		\end{equation}
		with $E_k=\sqrt{(\mu+t\cos k)^2+\Delta^2}$. The groundstate $|0\rangle$ is obtained from the condition $a_k |0\rangle=0$.
		
		The oscillating gate voltage induces an oscillating potential $V(\tau)=\delta\mu \cos(\omega \tau)$ described by
		\begin{equation}
			H_{\text{drive}}(\tau)=-V(\tau) \sum_k c^\dagger_k c_k= \frac{1}{2} \sum_k \frac{V(\tau)}{E_k} \,(a^\dagger_{k}, a_{-k})\left(
			\begin{array}{cc}
				-\mu-t \cos k  & i \Delta \sin k \\
				-i \Delta \sin k &\mu+t \cos k \\
			\end{array}
			\right)\left( \begin{array}{c}
				a_{k}\\
				a^\dagger_{-k}
			\end{array}\right).
		\end{equation} 
		Here we switch to the Floquet formalism \cite{Rudner2020}. In the Heisenberg picture, we make the following Floquet-Bogoliubov ansatz for new Floquet quasi-particle operators, 
		\begin{align}
			\tilde a_k^\dagger(\tau)=\sum_{n=-N}^{N} u_{k,n} \, e^{i (E_k^\text{FL}+n \omega )\tau}\,  a_k^\dagger+ v_{k,n}\, e^{-i (E_k^\text{FL}+n \omega )\tau} \,a_{-k}.
		\end{align} 
		where $E_k^\text{FL}$ are the Floquet quasi-energies, defined modulo $\omega$. We choose the Floquet energies to be in the interval $0\le E_k^\text{FL} <\omega$ and $N$ determines the number of Floquet modes.
		For an exact solution, one has to take the limit $N\to \infty$, but for small $\delta \mu$, the result converges rapidly even for small $N$.
		The $u_{k,n}$ and $v_{k,n}$ are determined from  the condition (i) that the new operators obey the canonical anti-commutation relations of fermions and (ii) that $\tilde a_k^\dagger$ and $\tilde a_{-k}$ fulfill the Heisenberg equations of motion. This results in a  matrix equation, $E_k^\text{FL} w^\text{FL} = H_k^\text{FL} w^\text{FL}$ where $H_k^\text{FL}$ is the (single-particle) Floquet Hamilitonian (or ``Floquet matrix'') for fixed momentum $k$ and the vector $w^\text{FL}$ has the $2 (2 N+1)$ components  $u_{k,n}$ and $v_{k,n}$, $-N\le n\le N$.
		Importantly, these conditions do {\em not} completely fix  the quasi-particle operators as one can always perform a particle-hole transformation (not to be confused with a particle-hole symmetry operation \cite{Zirnbauer2021})
		\begin{align}
			\tilde a_k^\dagger \longleftrightarrow \tilde{a}_{-k}, \qquad E_k^\text{FL}  \longleftrightarrow - E_k^\text{FL}+ \omega.\label{eq:ambiguity}
		\end{align}
		We added $+\omega$ in the last equation to ensure that the quasi-energy remains in the Floquet zone, $0\le -E_k^\text{FL}+\omega<\omega$. In a more general setup where the $k \to -k$ symmetry is absent, one can, equivalently, consider the discrete transformation $\tilde a_m^\dagger \leftrightarrow \tilde a_m$ and 
		$E_m^\text{FL}  \longleftrightarrow - E_m^\text{FL}+ \omega$ for Floquet-Bogoliubov states with quantum number $m$.
		
		The ambiguity of Eq.~\eqref{eq:ambiguity} can be fixed by observing that the annihilation operators have to fulfill an extra condition: they should annihilate
		the adiabatically prepared initial state $|\psi_0\rangle$,
		\begin{align}
			\tilde a_k |\psi_0 \rangle =0.
		\end{align}	
		Thus the preparation protocol of the state $|\psi_0\rangle$ is needed to identify correct annihilation and creation operators and to resolve the discrete ambiguity in the operators and quasi-energies expressed in Eq.~\eqref{eq:ambiguity}. 
		For any adiabatic preparation protocol there is a simple and unique way to identify the correct ground state and thus the correct quasi  energies of excitations, $\tilde a_k |\psi_0 \rangle$. The starting point is the groundstate of a static Hamiltonian, where all (physical) excitation energies are by definition positive. Then we can simply use the principle of adiabatic continuity to track the operators: a creation operator stays a creation operator during adiabatic evolution. 
		
		As we have shown in the main text, different adiabatic protocols (frequency sweep vs. amplitude sweep) with the same final $H(\tau)=H_{\text{BdG}}+H_\text{drive}(\tau)$ lead to different sets of creation and annihilation operators and thus different physical excitation energies.

		We have done this adiabatic tagging of operators numerically for the system with open boundary conditions simply by tracking the evolution of excitation energies during the adiabatic evolution. Below, we give an analytically tractable example
		by considering weak oscillations, $\delta \mu \ll \omega,\Delta$, in an infinite system. 
		In this case one can focus on approximately resonant processes with $E_k \approx -E_k+\omega$. Ignoring all non-resonant processes, the infinitely large Floquet matrix can be reduced to a simple $2 \times 2$ matrix,
		\begin{equation}
			H^{\text{FL}}_k=\left(
			\begin{array}{cc}
				-E_k +\omega & \delta\mu \,f_k \\
				\delta\mu\, f^*_k & E_k \\
			\end{array}
			\right)
		\end{equation}
		with $f_k= -i\frac{\Delta \sin k}{ 2 \sqrt{(\mu+t\cos k)^2+\Delta^2 \sin^2 k}}$. 
		The eigenvalues of this $2\times 2$ matrix are thus $\frac{\omega}{2}\pm \sqrt{\left(\delta\mu |f_k|\right)^2+\left(E_k-\omega/2\right)^2 }$ where one of the energies is a physical excitation quasi-energy (multiplying 
		$\tilde a_k^\dagger \tilde a_k$ in the second-quantized formula) while the other quasi-energy is the Bogoliubov shadow (multiplying 
		$\tilde a_k \tilde a_k^\dagger$). Which one is which, depends on the protocol.
		
		Let us consider the protocol where $\delta \mu$ is increased  adiabatically at fixed frequency $\omega$. We track the energies back to $\delta \mu =0$ and demand that the physical excitation quasi-energy matches the physical excitation energy $E_k$ of the initial state
		\begin{align}
			\lim_{\delta \mu \to 0} \frac{\omega}{2}\pm \sqrt{(\delta\mu |f_k|)^2+(E_k-\omega/2)^2 }= \frac{\omega}{2}\pm |E_k-\omega/2 | \stackrel{!}{=} E_k.
		\end{align}
		Thus we should use the $+$ sign  ($-$ sign) for $E_k>\omega/2$ ($E_k<\omega/2$). The 
		physical excitation quasi-energies are then given by \begin{align}
			E^\text{FL}_k=\left\{ \begin{array}{ll}
				\frac{\omega}{2} + \sqrt{(\delta\mu |f_k|)^2+(E_k-\omega/2)^2 } &\text{for } E_k>\omega/2 \\[1mm]
				\frac{\omega}{2} - \sqrt{(\delta\mu |f_k|)^2+(E_k-\omega/2)^2 } &\text{for } E_k<\omega/2 
			\end{array}
			\right. .\label{eq:prot1}
		\end{align}
		There is therefore a jump in the excitation quasi-energy at $k=k_0$ where $E_{k_0}-\omega/2$ changes sign, see Fig.~\ref{fig:jump} of the main text.

		Let us consider the``frequency-sweep'' protocol, see Fig.~\ref{fig:protocol} of the main text.
		Here one starts by increasing $\delta \mu$ using a small frequency where $E_k>\omega/2$ for all $k$. Thus, the physical excitation quasi-energy always has the $+$ sign in front of the square root. If one increases in a second step $\omega$ to reach the same final state as above, one always stays in the $+$ branch. Thus the physical excitation energies, in this case, are simply 
		\begin{align}
			E^\text{FL}_k=
			\frac{\omega}{2} + \sqrt{(\delta\mu |f_k|)^2+(E_k-\omega/2)^2 } \label{eq:prot2}
		\end{align}
		for all momenta $k$. We would like to stress that the different excitation energies of the two protocols, Eqs.~\eqref{eq:prot1} and \eqref{eq:prot2}, arise because two very different many-particle wave functions are created in the two cases.
		
	\end{widetext}
\end{document}